\documentclass[doublespacing]{elsart}

\usepackage{icarus}

\usepackage{natbib}
\usepackage{aas_macros}
\usepackage{array}

\usepackage{hyperref}
\usepackage{csquotes}
\hypersetup{colorlinks=true,linkcolor=blue,citecolor=blue,urlcolor=blue}

\bibpunct{(}{)}{;}{a}{,}{,}

\usepackage{graphicx}

\begin{document}

\begin{frontmatter}


\title{3D modeling of organic haze in Pluto's atmosphere}


\author[a]{Tanguy Bertrand}, 
\author[a]{Fran\c cois Forget},

\address[a]{Laboratoire de M\'et\'orologie Dynamique, CNRS/UPMC (France)}

\begin{center}
\scriptsize
Copyright \copyright\ 2005, 2006 Ross A. Beyer, David P. O'Brien, Paul
Withers, and Gwen Bart
\end{center}


%
%
%
%
%


\end{frontmatter}



\begin{flushleft}
\vspace{1cm}
Number of pages: \pageref{lastpage} \\
Number of tables: \ref{lasttable}\\
Number of figures: \ref{lastfig}\\
\end{flushleft}


\begin{pagetwo}{Haze on Pluto}

Tanguy Bertrand \\
Laboratoire de M\'et\'orologie Dynamique, CNRS/UPMC (France)\\

Email: tanguy.bertrand@lmd.jussieu.fr\\

\end{pagetwo}

\begin{abstract}

The New Horizons spacecraft, which flew by Pluto on July 14, 2015, revealed the presence of haze in Pluto's atmosphere that were formed by CH$_{4}$/N$_{2}$ photochemistry at high altitudes in Pluto's atmosphere, as on Titan and Triton. In order to help the analysis of the observations and further investigate the formation of organic haze and its evolution at global scales, we have implemented a simple parametrization of the formation of organic haze in our Pluto General Circulation Model. The production of haze in our model is based on the different steps of aerosol formation as understood on Titan and Triton: photolysis of CH$_{4}$ in the upper atmosphere by Lyman-$\alpha$ UV radiation, production of various gaseous species, and conversion into solid particles through accumulation and aggregation processes. 
The simulations use properties of aerosols similar to those observed in the detached haze layer on Titan.
We compared two reference simulations ran with a particle radius of 50 nm: with, and without South Pole N$_{2}$ condensation. We discuss the impact of the particle radius and the lifetime of the precursors on the haze distribution. We simulate CH$_{4}$ photolysis and the haze formation up to 600 km above the surface. 
Results show that CH$_{4}$ photolysis in Pluto's atmosphere in 2015 occured mostly in the sunlit summer hemisphere with a peak at an altitude of 250 km, though the interplanetary source of Lyman-$\alpha$ flux can induce some photolysis even in the Winter hemisphere. We obtained an extensive haze up to altitudes comparable with the observations, and with non-negligible densities up to 500 km altitude. 
In both reference simulations, the haze density is not strongly impacted by the meridional circulation. With no South Pole N$_{2}$ condensation, the maximum nadir opacity and haze extent is obtained at the North Pole. With South Pole N$_{2}$ condensation, the descending parcel of air above the South Pole leads to a latitudinally more homogeneous haze density with a slight density peak at the South Pole.
The visible opacities obtained from the computed mass of haze, which is about 2-4$\times10^{-7}$ g\,cm$^{-2}$ in the summer hemisphere, are similar for most of the simulation cases and in the range of 0.001-0.01, which is consistent with recent observations of Pluto and their interpretation.

\end{abstract}

\begin{keyword}
Pluto\sep Atmosphere\sep Haze\sep Modeling\sep GCM\sep \\
\texttt{http://icarus.cornell.edu/information/keywords.html}
\end{keyword}

\section{Introduction}

Pluto, Titan and Triton all have a nitrogen-based atmosphere containing a significant fraction of methane, an efficient recipe known to lead to the formation of organic haze in the atmosphere, as confirmed by observations \citep{Toma:05,RagePoll:92,HerbSand:91,Ster:15} and laboratory experiments \citep{Trai:06,Rann:10,Lavv:08}. Here, we use the Global Climate Model of Pluto (herein referred to as GCM), developed at the Laboratoire de M\'et\'eorologie Dynamique (LMD) and designed to simulate the atmospheric circulation and the methane cycle on Pluto and to investigate several aspects of the presence of haze at a global scale on Pluto \citep{Forg:16,Bert:16}.
What controls haze formation on Pluto? At which altitudes and latitudes does it form and where does sedimentation occur? What amount of particles forms the haze, and what is its opacity? To address those key questions we have developed a simple parametrization of haze in the GCM. The parametrization is based on a function of aerosols production, which directly depends on the amount of the Lyman-$\alpha$ UV flux. The photolysis reaction of CH$_{4}$ is photon-limited. That is, all incident photons are absorbed by the CH$_{4}$ molecules present in Pluto's atmosphere. 

During the flyby of the Pluto system on July 14, 2015, the New Horizons spacecraft recorded data about the structure, composition and variability of Pluto's atmosphere. In particular, Alice, the UV spectrometer on-board, observed solar occultations of Pluto's atmosphere which help to determine the vertical profiles of the densities of the present atmospheric constituents and provide key information about the haze. Within this context, our work aims to help the analysis of the New Horizons observations with model predictions of the possible evolution, spatial distribution and opacity of haze in Pluto's atmosphere and on its surface. 


We begin in Section \ref{background} with a background on haze formation processes as understood on Titan, Triton and Pluto. In Section \ref{model} we describe the GCM.
The parametrization of organic haze, as well as its implementation in the model are described step by step in Section \ref{param}. Finally, results are shown in Section \ref{results} for two climate scenarios: with and without South Pole N$_{2}$ condensation.

\section{Background on planetary haze formation}
\label{background}

One of Titan's most fascinating features is the dense and widespread organic haze shrouding its surface and containing a large variety of molecules which strongly impact the global climate. This makes Titan a perfect place to study organic chemistry and the mechanisms involved in a planetary haze formation. 
Since 2004, the exploration of Titan's haze by the Cassini/Huygens mission has provided a large amount of observational data, revealing complex chemistry, particularly at high altitudes. This has stimulated more interest in understanding this phenomenon. 
The haze on Titan is vertically divided into two regions: a main haze up to 300 km altitude, and a thinner, overlying detached haze typically between 400-520 km \citep{Lavv:09}, whose origin is thought to be dynamic \citep{Rann:02}, although other scenarios were suggested \citep{Lars:15}. Both layers contain solid organic material resulting from photochemistry and microphysical mechanisms, some of which remain unknown \citep{Lebo:02, WilsAtre:03, Lavv:08}.

First, methane and nitrogen molecules are dissociated and ionized in the upper atmosphere (up to 1000 km above the surface) by solar UV radiation, cosmic rays and energetic electrons from Saturn's magnetosphere \citep{Sitt:10}. It is commonly thought that the molecules resulting from photolysis chemically react with each other, which leads to the formation of larger and heavier molecules and ions such as hydrocarbons, nitriles and oxygen-containing species \citep[e.g.]{Niem:10, Crav:06, Coat:07, Wait:07, Crar:09}. While CH$_{4}$ is easily destroyed by photolysis and provides most of the organic materials, N$_{2}$ is dissociated as well by extreme UV radiation which explains the rich composition of Titan's upper atmosphere. 
In particular, observations from Cassini and Huygens spacecrafts show the presence of hydrocarbons and nitriles, such as  C$_{2}$H$_{2}$, C$_{2}$H$_{4}$, C$_{2}$H$_{6}$, C$_{4}$H$_{2}$, C$_{6}$H$_{6}$, and HCN, as well as other more complex organics \citep{Shem:05}. These species, formed after photolysis in the upper atmosphere, are the precursors of the haze.  
Then, through multiple processes of sedimentation, accumulation and aggregation, the precursors are thought to turn into solid organic aerosols which become heavy enough to form the orange haze surrounding the moon as seen in visible wavelengths \citep{West:91, Rann:95, Yell:06, Lavv:09}. These aerosols are thought to be aggregates (modeled as fractal-like particles) composed of many spherical particles (monomers) that bond to each other. On Titan, the aerosols start to become large enough to be visible in the detached haze layer around 500 km altitude. Typically, they grow spherical up to radius 40-50 nm and then form fractal particles with monomer sizes of around 50~nm \citep{Lavv:09}.

What are the haze's dominant pathways? What are the chemical natures of complex haze particles? 

Several microphysical models \citep{Toon:92,Rann:97,Lavv:09} and photochemical models \citep{WilsAtre:04, Lavv:08, Hebr:13} have been developed, combining both transport and chemistry effects. The formation mechanisms of aerosol particles in Titan's atmosphere have also been investigated using laboratory experiments. By performing UV irradiation of CH$_{4}$ in a simulated Titan atmosphere, several experiments have been successful in producing solid particles and have found that they contain mostly high-molecular-weight organic species \citep[e.g.,][]{Khar:84,Khar:02,Coll:99, Iman:04, Szop:06,Gaut:12}. Experimental results from \citet{Trai:06} also show a linear relationship between the rate of aerosol production and the rate of CH$_{4}$ photolysis. In addition, they found that an increased CH$_{4}$ concentration could lead to a decrease in aerosol production in photon-limited reactions (this could be due to reactions between CH$_{4}$ and precursors forming non-aerosol products). 

Titan's atmosphere is not the unique place where organic haze can form. First, similar processes of haze formation are also thought to occur on Triton but yield less haze. During the Voyager 2 flyby in 1989, evidence of a thin haze was detected in Triton's atmosphere from limb images taken near closest approach \citep{Smit:89, Poll:90, RagePoll:92} and from Voyager 2 UVS solar occultation measurements \citep{HerbSand:91, Kras:92, Kras:93}. These data enabled the mapping of the horizontal and vertical distribution of CH$_{4}$ and haze as well as estimation of radiative and microphysical properties of the haze material. 
Analyses showed that the haze is present nearly everywhere on Triton, from the surface up to 30 km at least \citep{Poll:90}, where it reached the limit of detectability. Vertical optical depth derived from observations were found to be in the range 0.01-0.03 at UV wavelength 0.15~$\mu$m, and 0.001-0.01 at visible wavelength 0.47~$\mu$m. Haze particle sizes were estimated to be spherical and small, around 0.1-0.2~$\mu$m \citep{Kras:92, RagePoll:92, Poll:90}.
As on Titan, complex series of photochemical reactions may be involved in the formation of this haze, starting with CH$_{4}$ photolysis by the solar and the interstellar background Lyman-$\alpha$ radiation in the atmosphere of Triton at altitudes between 50-100 km, producing hydrocarbons such as C$_{2}$H$_{2}$, C$_{2}$H$_{4}$, C$_{2}$H$_{6}$ \citep{Stro:90, KrasCrui:95}. Dissociation of N$_{2}$ molecules is also suggested in the upper atmosphere around 200-500 km. Transitions between haze precursors to solid organic particles are still incompletely known, but it is commonly thought that it involves similar mechanisms to those on Titan.
Secondly, organic chemistry has also been studied in the Early Earth climate context, where a scenario of a N$_{2}$/CH$_{4}$ atmosphere is plausible to form a hydrocarbon haze \citep{Trai:06}.

\begin{table}[h]
\begin{center}
\begin{tiny} 
\begin{tabular}{>{\bfseries}llll}
 & \textbf{Titan} & \textbf{Triton} & \textbf{Pluto (2015)} \\
\hline
Distance from Sun (UA) & 9.5 & 30 & 32.91 \\
Solar Flux (ph\,m$^{-2}$\,s$^{-1}$) & $4.43\times10^{13}$ & $4.44\times10^{12}$ & $3.69\times10^{12}$\\
CH$_{4}$ mixing ratio & $1.5\%^{a}$ & $0.02\%^{b}$ & $0.6\%^{c}$\\
CO mixing ratio & $0.0045\%$ & $0.07\%^{b}$ & $0.05\%^{c}$\\
P$_{est}$ (kg\,m$^{-2}$\,s$^{-1}$) & $2.94\times10^{-13}$ & $7.47\times10^{-14}$ & $5.98\times10^{-14}$\\
P$_{lit}$ (kg\,m$^{-2}$\,s$^{-1}$) & $0.5-3\times10^{-13}$ $^{d}$ & $6.0\times10^{-14}$ $^{e}$ &  $9.8\times10^{-14}$ $^{f}$ \\
\hline
\multicolumn{4}{l}{$^{a}$\textit{above the tropopause, \citet{Niem:10}}}  \\
\multicolumn{4}{l}{$^{b}$\textit{\citet{Lell:10}}} \\
\multicolumn{4}{l}{$^{c}$\textit{\citet{Lell:11}}} \\
\multicolumn{4}{l}{$^{d}$\textit{\citet{WilsAtre:03,McKa:01}}} \\
\multicolumn{4}{l}{$^{e}$\textit{\citet{StroSumm:95}}} \\
\multicolumn{4}{l}{$^{f}$\textit{\citet{Glad:16}}} \\
\hline   
\end{tabular}
\end{tiny}
\caption{Comparison of the incident UV flux and fraction of methane for a first order estimation of aerosol production rates on Titan, Triton and Pluto. The estimated rate P$_{est}$ is compared to the observed rate P$_{lit}$, as detailed in the literature.\newline}
\label{Table1}
\end{center}
\end{table}

\begin{table}[h]
\begin{center}
\begin{tiny}
\begin{tabular}{>{\bfseries}llll}
 & \textbf{Titan (at 400km)} & \textbf{Triton} & \textbf{Pluto}\\
\hline
Gravity (m$^{2}$\,s$^{-2}$) & 1.01 & 0.779 & 0.62\\
Pressure (Pa) & 1.5 & 1.4-1.9 & 1-1.1$^{a}$\\
Visible normal opacity & 0.07$^{b}$ & 0.003-0.008$^{c}$ & 0.004$^{a}$\\
\hline 
\multicolumn{4}{l}{$^{a}$\textit{\citet{Ster:15}}} \\
\multicolumn{4}{l}{$^{b}$\textit{\citet{Cour:11}}} \\
\multicolumn{4}{l}{$^{c}$\textit{\citet{RagePoll:92,Kras:92}}} \\
\hline   
\end{tabular}
\caption{Gravity,surface pressure and visible aerosol opacity on Pluto and Triton, compared to the the values encountered in the detached haze layer on Titan \newline}
\label{Table2}
\end{tiny}
\end{center}
\end{table}

Finally, the presence of a haze on Pluto was suspected \citep{Elli:89,Stan:89,Forg:14} and confirmed in 2015 by New Horizons. 

At high phase angles, Pluto's atmosphere revealed an extensive haze reaching up to 200 km above the surface, composed of several layers \citep{Ster:15}. Observations show that the haze is not brightest to the sub solar latitude, where the incoming solar flux is stronger, but to Pluto North Pole. 
The haze is strongly forward scattering in the visible with a blue color, while at the same time there is haze extinction optical depth exceeding unity in the UV. The blue color and UV extinction are consistent with a small size of about 10 nm for monomers, whereas the high forward scatter to back scatter ratio in the visible suggests a much larger overall size of at least 200 nm. Although the haze may contain particles of diverse sizes and shapes depending on the altitude, these properties may also be consistent with fractal aggregate particles composed of 10 nm monomers \citep{Glad:16,Chen:16}. 

Although the specific mechanisms of haze formation are not fully understood, it seems that the main parameters controlling the formation of haze in a N$_{2}$/CH$_{4}$ atmosphere are the fractional amount of CH$_{4}$ (enough CH$_{4}$ is required to avoid CH$_{4}$-limited reactions, that is when the CH$_{4}$ concentration in the atmosphere is not sufficient to absorb all incoming photons) and the UV flux available to photolyze it. 

One can compare the UV flux and the fraction of methane for Titan, Triton and Pluto to estimate the haze formation rate to first order. 
Here we assume that the impact of cosmic rays and energetic electrons from Saturn's magnetosphere is negligible for this first order comparison.  
As shown on \autoref{Table1} and \autoref{Table2}, Pluto's atmosphere contains 10 times less CH$_{4}$ and receives 10 times less solar UV flux than Titan (relative to the atmospheric mass). Consequently, it is likely that CH$_{4}$ photolysis on Pluto leads to the formation of haze aerosols (and precursors) in lower quantities than on Titan. Compared to Triton, Pluto has similar surface pressure and gravity and its atmosphere contains 10 times more CH$_{4}$, for a comparable UV flux. Thus, similar amounts of haze are expected on Pluto and Triton, depending on the accelerating or decelerating role of larger CH$_{4}$ amount. \citet{Ster:15} reported a visible normal opacity of 0.004 on Pluto, which is in the range of what has been observed on Triton, although it also depends on the scattering properties of haze particles. 
On Titan, the pressure corresponding to the location of the detached haze layer at about 400 km altitude is about 1 Pa, which is similar to the surface pressure on Pluto in 2015. 
While \citet{Rann:03} predicted the peak of production of haze in Titan's GCMs at a pressure around 1.5 Pa, Cassini observations \citep{Wait:05, Tean:12} pointed to active chemistry and haze formation at lower pressures. In addition, the amounts of methane at these altitudes on Titan and in Pluto's atmosphere are of the same order of magnitude. Thus, Pluto has sufficient pressure and material in its atmosphere so that complex and opaque organic aerosols form, in a manner similar to the detached haze layer on Titan.
Consequently, in this paper, we use the microphysical and single scattering optical properties of Titan detached haze around 400 km altitude as a reference to define the haze properties on Pluto while the mass of aerosols is calculated by the model without any empirical assumption.

\section{Model description}
\label{model}

The LMD Pluto General Circulation Model (GCM) contains a 3D Hydrodynamical core inherited and adapted from the LMD Mars GCM \citep{Forg:99}. It is described in more details in \citet{Forg:16}. The large-scale atmospheric transport is computed through a "grid point model"
composed of 32 longitude and 24 latitude points. A key difference with the \citet{Forg:16} version of the model is that we use 28 layers instead of 25 to extend the model top up to about 600 km, with most of the layers in the first 15 km in order to obtain a finer near-surface resolution, in the boundary layer. The horizontal resolution at the equator is typically around 170 km. 
The physical part of the model, which forces the dynamics, takes into account the N$_{2}$ and the CH$_{4}$ cycles (condensation and sublimation in both the atmosphere and the ground), the vertical turbulent mixing and the convection in the planetary boundary layer, the radiative effect of CH$_{4}$ and CO, using the correlated-k method to perform a radiative transfer run and taking into account NLTE effects, a surface and subsurface thermal conduction model with 22 layers and the molecular conduction and viscosity in the atmosphere. 

\section{Modeling haze on Pluto}
\label{param}

Here we describe our representation of the organic haze formation and transport in the GCM. 
The driving force of the photochemical reactions occurring in a N$_{2}$-CH$_{4}$ atmospheric layer is the UV flux received by this layer. First we consider the photolysis of CH$_{4}$ by Lyman-$\alpha$ only (Section \ref{photolysis}), using the results from \citet{Glad:15} to calculate the incident Lyman-$\alpha$ flux at Pluto (Section \ref{lymanalpha}). We assume that each incident photon ultimately interacts with one molecule of methane, to form by photolysis haze precursors which can be transported by the circulation (Section \ref{mechanism}). Finally we convert haze precursors into organic haze using a constant characteristic decay time (Section \ref{conversion}). Haze particles properties used in this study are detailed in Section \ref{hazeproperties}. In order to validate this approach, we estimate the total aerosol production thus obtained on Pluto, Titan and Triton and compare with literature values in Section \ref{hazeproduction}. 

\subsection{Photolysis of CH$_{4}$ by Lyman-$\alpha$}
\label{photolysis}

We consider only the photolysis of CH$_{4}$  by the Lyman-$\alpha$ component of the UV spectrum. This is because the Hydrogen Lyman-$\alpha$ line at 121.6 nm is the strongest ultraviolet emission line in the UV solar spectrum where absorption by CH$_{4}$ happens. In fact, the solar irradiance between 0 and 160 nm (far ultraviolet) is dominated by the Lyman-$\alpha$ emission by a factor of 100. The UV solar irradiance grows significantly at wavelengths values higher than 200 nm (middle and near-ultraviolet) but N$_{2}$, CH$_{4}$ and CO do not absorb at these wavelengths.
Both N$_{2}$ and CH$_{4}$ absorb with similar efficiency in the UV but not at the same wavelengths. N$_{2}$ is the primary absorber at wavelength between 10 and 100 nm, while CH$_{4}$ absorbs mainly between 100 and 145 nm. Thus the interaction between CH$_{4}$ and Lyman-$\alpha$ emission dominates the other interactions between the UV flux and the N$_{2}$-CH$_{4}$ atmosphere by a factor of 100.
On Pluto, CO may also contribute to the formation of haze. It absorbs in the far UV spectrum at similar rates that N$_{2}$. However, at 121.6 nm, it absorbs 10 times less than CH$_{4}$. Here we chose to neglect the effect of N$_{2}$ and CO absorption.
This first assumption enables us to write Beer's law as the following:

\begin{equation}
I(\lambda,P) = I_{0} \, e^{-\int_{0}^{P}\frac{\sigma_{CH4}\,N_{a}\,q_{CH4}}{M_{CH4}\,g}\,\frac{dP}{cos(\theta)}}
\label{Beerlaw}
\end{equation}

where $I_{0}$ is the incident intensity (in ph\,m$^{-2}$\,s$^{-1}$) and $I(\lambda,P)$ the intensity after absorption for a given wavelength $\lambda$ and pressure $P$, $\sigma_{CH4}$ is the absorption cross section of CH$_{4}$ at wavelength $\lambda$ (here in m$^{2}$\,molec$^{-1}$ but usually given in cm$^{2}$\,molec$^{-1}$), $q_{CH4}$ is the mass mixing ratio of CH$_{4}$ at pressure $P$ (kg\,kg$^{-1}_{air}$), $M_{CH4}$ is the methane molecular mass (kg\,mol$^{-1}$), $N_{a}$ is the Avogadro constant, $\theta$ is the flux incident angle and g the surface gravity. We use $\sigma_{CH4}=1.85\times10^{-17}$ cm$^{2}$ at Lyman-$\alpha$ wavelength \citep{Kras:04} and $q_{CH4}$ as calculated by the GCM for each vertical layer. The calculation of the Lyman-$\alpha$ flux radiative transfer is performed independently for the solar and the interplanetary medium fluxes in order to take into account different values for the incident flux $I_{0}$ and the incident angle $\theta$ (see Section \ref{lymanalpha}).

\subsection{Sources of Lyman-$\alpha$}
\label{lymanalpha}

The sources of Lyman-$\alpha$ flux at Pluto are adopted from \citet{Glad:15}, which takes into account the solar as well as the interplanetary medium (IPM) Lyman-$\alpha$ fluxes. The IPM emission corresponds to interplanetary hydrogen atoms passing through the solar system which resonantly scatter solar Lyman-$\alpha$ photons and thus diffuse Lyman-$\alpha$ emission. Therefore the total Lyman-$\alpha$ flux at any pressure level $P$ in Pluto's atmosphere is:

\begin{equation}
I_{tot}(P) = I_{sol}(P) + I_{IPM}(P)
\label{Itot}
\end{equation}

The solar Lyman-$\alpha$ flux at Pluto is inversely proportional to the square of the Sun-Pluto distance. It is obtained by considering a constant solar Lyman-$\alpha$ flux at Earth of $4\times10^{15}$ ph\,m$^{-2}$\,s$^{-1}$ and a constant extinction factor of 0.875 due to the interaction with interplanetary hydrogen between Pluto and the Sun, which are values estimated by \citep{Glad:15} for 2015. The solar Lyman-$\alpha$ flux $I^{sol}_{0}$ thus estimated at Pluto is $3.23\times10^{12}$ ph\,m$^{-2}$\,s$^{-1}$. The incident angle $\theta^{sol}$ corresponds to the solar zenith angle. 

\begin{figure}[h]
\begin{center}
\includegraphics[width=6in]{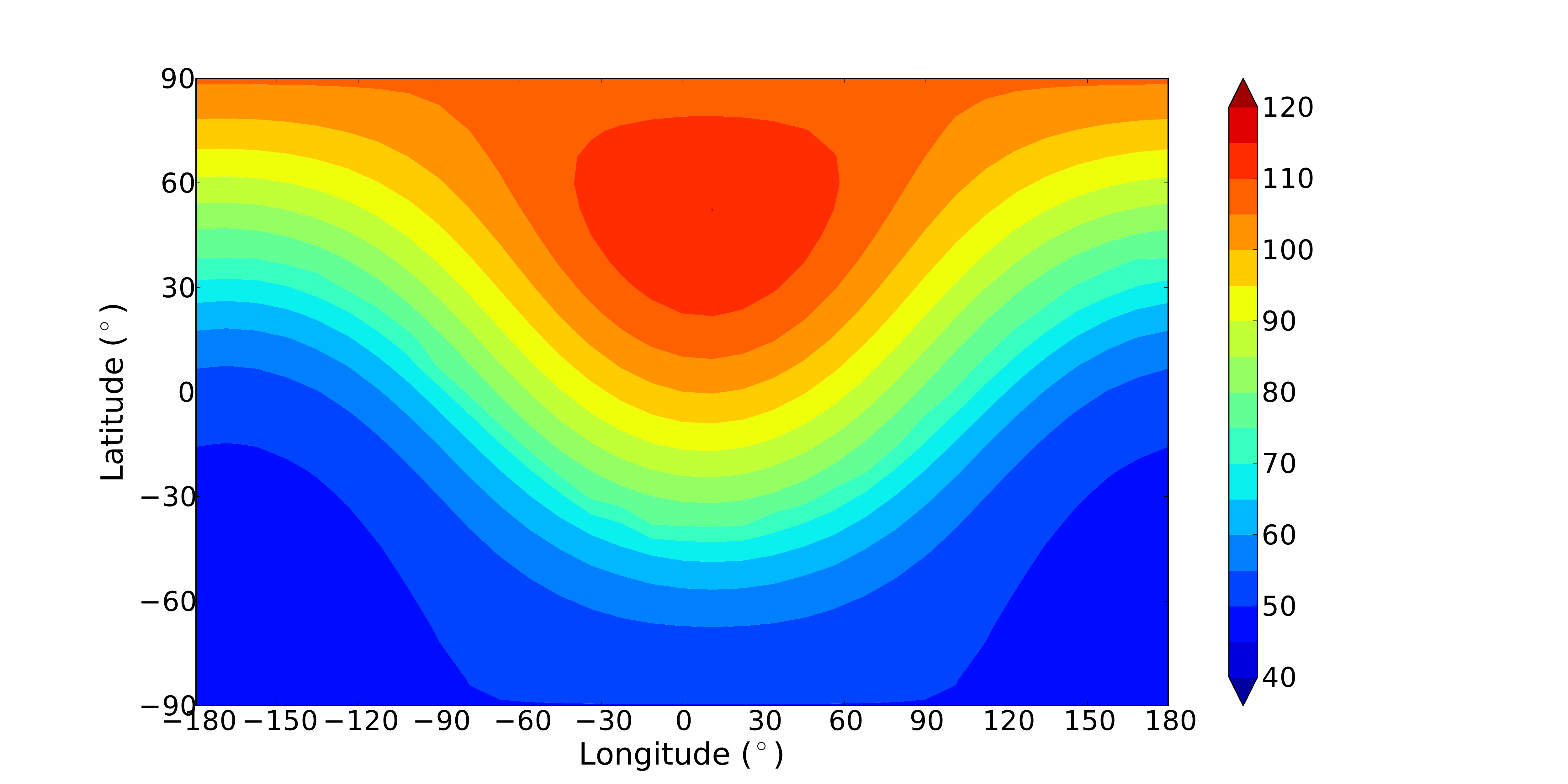}
\caption{An instantaneous map of interplanetary Lyman-$\alpha$ emission (10$^{10}$ ph\,m$^{-2}$\,s$^{-1}$) on Pluto in July 2015, estimated by integrating the all-sky IPM brightness given by figure 4 in \citet{Glad:15} over the half celestial sphere at each point of the map. In this example, the subsolar longitude is the sub Charon longitude (0\char23) \newline}
\label{fit}
\end{center}
\end{figure}

The IPM Lyman-$\alpha$ source at Pluto is not isotropic, as shown on figure 4 in \citet{Glad:15}, which presents the all-sky brightness of IPM emissions at Pluto in Rayleigh units in 2015. The brightness is stronger near the subsolar point and is minimal in the anti-sunward hemisphere. In order to take into account this property in the parametrization and compute the number of photons entering Pluto's atmosphere at a given location, we integrated the all-sky IPM brightness estimated in 2015 from \citet{Glad:15} over the half celestial sphere as seen at the considered location. The flux $I^{IPM}_{0}$ obtained varies with the local time but does not strongly depend on the Sun-Pluto distance (we use the flux estimated in 2015 for all other years). \autoref{fit} shows the final result: we find a maximum flux at subsolar point of 1.15$\times10^{12}$ ph\,m$^{-2}$\,s$^{-1}$, a minimum flux at anti-subsolar point of 4.90$\times10^{11}$ ph\,m$^{-2}$\,s$^{-1}$ and an average flux over the planet of 7.25$\times10^{11}$ ph\,m$^{-2}$\,s$^{-1}$. 
We consider that the incident angle for the IPM flux $\theta^{IPM}$ is equal to the solar zenith angle during daytime, when the IPM flux is dominated by the forward scattered halo of the solar flux. When the solar zenith angle is greater than $\pi$/3 (nighttime), we consider that the IPM flux is more isotropic and we set the incident angle to $\pi$/3.

At the Sun-Pluto distance during New Horizon flyby (32.91 UA), this IPM source of Lyman-$\alpha$ is significant compared to the solar source. Considering the solar Lyman-$\alpha$ flux, the energy of a photon at Lyman-$\alpha$ wavelength (121.6 nm) and its dissipation over the whole surface of Pluto (the initial flux is divided by a factor of 4), the power of solar Lyman-$\alpha$ source at Pluto obtained is 22.93 MW. The same calculation can be performed for the IPM flux. \citet{Glad:15} gives an averaged IPM brightness at Pluto of 145 R (1 R = 1/ $4\pi$ $\times10^{10}$ ph\,m$^{-2}$\,s$^{-1}$\,sr$^{-1}$), which corresponds to a flux of $1.45\times10^{12}$ ph\,m$^{-2}$\,s$^{-1}$ once integrated on the celestial sphere. This leads to a contribution of IPM Lyman-$\alpha$ source at Pluto of 10.30 MW. Consequently, solar and IPM sources at Pluto account for respectively $70\%$ and $30\%$ of the total power source. 

\newpage

\subsection{Production of haze precursors}
\label{mechanism}

In the parametrization, we consider that each absorbed Lyman-$\alpha$ photon destroys one molecule of methane by photolysis, thus forming haze precursors (CH$_{3}$, CH$_{2}$ , CH + N, etc.) converted later into aerosols. Using equation \ref{Beerlaw} and \ref{Itot}, the precursors production rate (in kg\,kg$^{-1}_{air}$\,s$^{-1}$) is calculated as: 

\begin{equation}
P_{prec}(P) = \frac{M_{CH4}\,g}{N_{a}}\,\frac{dI_{tot}}{dP}
\label{procprod}
\end{equation}

In the model, all possible precursors which can form during this reaction are represented by a unique gas. The equation of the reactions is: 

\begin{equation}
CH_{4} + h\nu \rightarrow precursors \rightarrow haze\; aerosols
\label{reaction}
\end{equation}

This mechanisms correlates linearly the rate of haze precursors production with the rate of CH$_{4}$ photolysis. It has also been used by \citet{Trai:06} to estimate aerosols production on Titan and Early Earth. 
In reality, the reactions are more complex and could lead to the irreversible production of HCN, or to the production of molecules such as C$_{2}$H$_{2}$ or C$_{2}$H$_{6}$ which can later be photolyzed themselves as well. In addition, CH$_{4}$ molecules may be chemically dissociated by reacting directly with the precursors. Consequently, these reactions could lead either to an increase in the amount of carbon atoms available as haze material, increasing the haze production, or to non-aerosol products, slowing down the haze production \citep{Trai:06}.

In the parametrization, the haze production is regulated by a factor $K_{CH4}$, that corresponds to the ratio between the total number of carbon atoms in the tholins and the number of carbon atoms coming from CH$_{4}$ photolysis. $K_{CH4}$ would range from 1 to 2 (respectively all or half of the carbon in the tholins are formed by direct CH$_{4}$ photolysis) if direct reactions between precursors and CH$_{4}$ occur and contribute to provide tholins with carbon atoms. However, the ratio could be lower than 1 considering the formation of other non-aerosol products (see Section \ref{sec:sensib_mass}).

Additionally, nitrogen may contribute to the chemical reactions and provide material for aerosol formation. In order to take into account this process, the haze production is also boosted by a factor $K_{N}$=1+$N/C$, $N/C$ representing the mass ratio between nitrogen and carbon atoms contribution observed in the tholins (since molar masses of nitrogen and carbon are quite similar, the mass ratio is close to the number ratio). Different values of this ratio have been observed in laboratory experiments, ranging from 0.25 to 1 depending on the pressure (the higher the pressure, the lower the ratio), the temperature and the amount of methane in the simulated atmosphere \citep[e.g.][]{Coll:99,Tran:08,Nnam:13}. In the model, we adopt $N/C$ = 0.5, in line with the values obtained in \citet{Nnam:13} at low pressure, and $K_{CH4}$ = 1, so that the total production of tholins remains in the range of estimated values on Titan and Pluto (see Section \ref{hazeproduction}).

\subsection{Conversion of haze precursors to aerosols}
\label{conversion}

As the mechanisms at the origins of formation of organic haze are not well known, another assumption is made in the parametrization: we consider that the precursors become solid organic particles (by a set of processes of aggregation and polymerization that are not represented) after a given time. In practice, the amount of precursors is subject to exponential decay and is converted into aerosols with characteristic decay time $\tau$ (or characteristic time for aerosol growth).
In other words, $\tau$ is the mean lifetime of the precursors before they become solid aerosols. This time is difficult to estimate as it depends on atmospheric conditions (concentration, pressure...). However, Titan's atmospheric models show that the time needed for precursors to evolve from the photolysis area to the detached layer is typically around 10$^{6}$-10$^{8}$ s \citep{Lavv:11,Rann:93}. 
Consequently, we used in our reference GCM simulations a value of 10$^{7}$ s for Pluto aerosols and we examine the sensitivity of the results to this parameter in Section \ref{sec:sensib_tau}.

Once produced, the aerosols are transported by the atmospheric circulation, mixed by turbulence, and subject to gravitational sedimentation (see Section \ref{hazeproperties}).

\subsection{Discussion on total aerosol production}
\label{hazeproduction}

Equation \ref{reaction} enables us to estimate the total haze production rate $P$ (kg\,m$^{-2}$\,s$^{-1}$) in a N$_{2}$/CH$_{4}$ atmosphere:

\begin{equation}
P=(F_{SOL}+F_{IPM})\,\frac{M_{CH4}}{N_{a}}\,K_{CH4}\,K_{N} \quad  with  \quad
F_{SOL}=\frac{I_{Earth}}{4\,{d_{P}}^{2}}\,E_{H}
\label{production}
\end{equation}

where F$_{SOL}$ and F$_{IPM}$ are the solar and IPM Lyman-$\alpha$ flux respectively (in ph\,m$^{-2}$\,s$^{-1}$), $M_{CH4}$ is the molar mass of methane ($M_{CH4}=16\times10^{-3}$ kg\,mol$^{-1}$), $N_{a}$ is the Avogadro constant, I$_{Earth}$ is the initial Lyman-$\alpha$ flux at Earth (we set $I_{Earth}$=$4\times10^{15}$ ph\,m$^{-2}$\,s$^{-1}$), d$_{P}$ is the distance in astronomical units of the considered planet $P$ to the Sun and E$_{H}$ is a constant extinction factor due to interaction with interplanetary hydrogen between the planet $P$ and the Sun. Here E$_{H}$ is set to 0.875 for the case of Pluto \citep{Glad:15} and to 1 for the other cases. The solar flux F$_{SOL}$ is equal to the incident solar flux $I^{sol}_{0}$ divided by a factor of 4 to take into account the distribution on the planetary sphere.  

It is important to note that the haze production rate is independent of the CH$_{4}$ concentration, even for CH$_{4}$ concentrations several orders of magnitude lower than on Pluto (see Section \ref{sec:sensib}). The reactions are photon-limited, i.e. that enough CH$_{4}$ is present in Pluto's atmosphere for all photons to be absorbed by CH$_{4}$. 

In order to validate the approach described by equation \ref{reaction}, we apply equation \ref{production} to Titan, Triton and Pluto and compare the haze production rates obtained with the literature. The values, obtained with $K_{CH4}$=1 and $K_{N}$=1.5, are summarized in \autoref{Table1}.
For Titan's case, we consider that the IPM flux is negligible compared to the solar flux. Using an average Sun-Titan distance $d_{Titan}$=9.5 UA, we find for Titan's atmosphere a Lyman-$\alpha$ flux of $1.11\times10^{13}$ ph\,m$^{-2}$\,s$^{-1}$ (dissipated on the planetary sphere) and a production rate of $2.94\times10^{-13}$ kg\,m$^{-2}$\,s$^{-1}$. This is comparable to values found by \citet{WilsAtre:03} and \citet{McKa:01}, as shown on \autoref{Table1}.
For Triton's case, we consider an averaged IPM flux of 340 R \citep{Broa:89,Kras:95}, which correspond to an IPM flux of $170\times10^{10}$ ph\,m$^{-2}$\,s$^{-1}$ distributed on the planetary sphere. Using an average Sun-Triton distance $d_{Titan}$=30 UA, we find for Triton's atmosphere a total Lyman-$\alpha$ flux (solar and IPM) of $2.81\times10^{12}$ ph\,m$^{-2}$\,s$^{-1}$ and a photolysis rate of $7.47\times10^{-14}$ kg\,m$^{-2}$\,s$^{-1}$, which is also in line with the literature references.
Since this approach provides good estimation of Titan's and Triton's total aerosol production, we used it to estimate the aerosol production rate for Pluto's atmosphere. \autoref{production} gives a production rate of $5.98\times10^{-14}$ kg\,m$^{-2}$\,s$^{-1}$ using the solar and IPM flux as calculated in Section \ref{lymanalpha}. 
This value is one order of magnitude lower than the one on Titan (due to the UV flux one order of magnitude lower) and comparable to the value found on Triton. It is of the same order of magnitude as the value estimated on Pluto from photochemical models \citep{Glad:16} shown in \autoref{Table1}.
 
\subsection{Properties of haze particles for sedimentation and opacity estimations}
\label{hazeproperties}

Haze precursors and particles are transported in the model by atmospheric circulation and are not radiatively active. In addition, the haze is considered too thin to affect the surface energy balance and does not change its ground albedo (in line with haze and surface observations on Triton as discussed in \citet{HillVeve:94}). 

The density of the aerosol material in the model is set to 800 kg\,m$^{-3}$, which is in the range of values typically used on Titan \citep{Soti:12,Lavv:13,Trai:06}.
The size of the haze particles affects their sedimentation velocity and thus the haze distribution in Pluto's atmosphere. In the GCM, we prescribe  a uniform size distribution of particles. For the reference simulations (with and without South Pole N$_{2}$ condensation), we assumed spherical particles with a radius of 50 nm, consistent with the properties of the detached haze layer on Titan (see Section \ref{background}). We also examine the sensitivity of the results to different sizes of particles in Section \ref{sec:sensib_r}, in order to bracket the different possible scenarios for Pluto's haze. 
We consider two lower radii of 30 nm and 10 nm, which is in the range of recent estimations \citep{Glad:16}, and one larger radius of 100 nm.

The particles fall with their Stokes velocity $\omega$, corrected for low pressures \citep{Ross:78}:

\begin{equation}
\omega=\frac{2}{9}\,\frac{r^{2}\,\rho\,g}{v}\,(1+\alpha\,Knud) \quad  with  \quad
Knud=\frac{k_{B}\,T}{\sqrt{2}\,\pi\,d^{2}\,p\,r}
\label{stokes}
\end{equation}

with $r$ the particle radius, $\rho$ the particle density, $g$ the Pluto's gravitational constant, $v$ the viscosity of the atmosphere, $Knud$ the Knudsen number, $p$ the considered pressure, $T$ the atmospheric temperature, $d$ the molecular diameter, $k_{B}$ the Boltzmann's constant and $\alpha$ a correction factor. 

On Pluto, the Knudsen number is significant and thus the sedimentation velocity is proportional to the particle radius. Consequently, in an ideal atmosphere without atmospheric circulation, a 100 nm particle will fall twice faster than a 50 nm particle, leading to a twice lower column mass of haze. Assuming an atmospheric temperature of 100 K and a surface pressure of 1 Pa, the sedimentation velocities above Pluto's surface are about 4.6$\times10^{-4}$, 1.4$\times10^{-3}$, 2.3$\times10^{-3}$ and 4.6$\times10^{-3}$ m\,s$^{-1}$ for an aerosol radius of 10, 30 50 and 100 nm respectively. 

One can note that the Stokes velocity is proportional to the inverse of the pressure. Theoretically, the lower the pressure, the higher the sedimentation velocity of the aerosol and thus the lower the mass of haze in the atmosphere. 

The choice of the size and the shape of aerosol particles is also critical to estimate their optical properties and thus their detectability. In Section \ref{sec:sensib_r}, we compare the opacities obtained with different particle radii. In Section \ref{opacity}, we examine the case of fractal particles by considering that they fall at the velocity of their monomers, due to their aggregate structure, which is only true for a fractal dimension equal to 2 \citep{Lavv:11,Lars:14}. 


\subsection{Description of the reference simulations}

In this paper, we compare two reference simulations which correpond to the two climate scenarios detailed in \citet{Forg:16}: 
One is the case of Sputnik Planum as the only reservoir of N$_{2}$ ice without N$_{2}$ condensation elsewhere (referred as No South Pole N$_{2}$ condensation), and the other is the case with a latitudinal band of N$_{2}$ ice at northern mid latitudes, as an additional reservoir of N$_{2}$ ice with Sputnik Planum, and an initially colder South Pole, allowing the N$_{2}$ ice to condense (with South Pole N$_{2}$ condensation).

The reference simulations study are defined as follows. A seasonal volatile model of Pluto is used to simulate the ice cycles over thousands of years and obtain consistent ices distribution, surface and subsurface temperatures as initial conditions for the GCM (see \citet{Bert:16} for more details). Then, GCM runs are performed from 1988 to 2015 included so that the atmosphere has time to reach equilibrium before 2015 (the spin up time of the model is typically 10-20 Earth years). The initial conditions, the settings of the model, as well as discussions about the sensitivity of the predictions to those settings can be found in \citet{Forg:16}.

The model is run with the haze parametrization using a precursor characteristic time for aerosol growth of 10$^{7}$ s (about 18 sols on Pluto), a fraction K$_{CH4}$=1 and K$_{N}$=1.5. The density and sedimentation effective radius of haze particles are set uniformly to 800 kg\,m$^{-3}$ and 50 nm respectively (see Section \ref{conversion}). \autoref{tab:inicond} summarizes the surface conditions and haze parameters used in the reference simulations \citep{Forg:16}. \newline

\begin{table}[h]
\begin{center}
\begin{tiny}
\begin{tabular}{llll}
\hline
\textbf{Global Thermal Inertia (J s$^{-0.5}$ m$^{-2}$ K$^{-1}$)} & 50 (diurnal) & 800 (seasonal) & \\
\textbf{Albedo} & 0.68 (N$_{2}$ ice) & 0.50 (CH$_{4}$ ice) & 0.15 (Tholins)\\
\textbf{Emissivity} & 0.85 (N$_{2}$ ice) & 0.85 (CH$_{4}$ ice) & 1 (Tholins)\\
\hline
\end{tabular}
\begin{tabular}{>{\bfseries}ll}
\hline 
Characteristic time for aerosol growth $\tau$ (s) & 10$^{7}$\\
K$_{CH4}$ & 1 \\
K$_{N}$ & 1.5 \\
Effective radius of haze particles (nm) & 50\\ 
Density of haze particles (kg.m$^{-3}$) & 800\\
\hline 
\end{tabular}
\caption{Surface conditions and settings for haze parametrization set for the GCM reference simulations \newline}
\label{tab:inicond}
\end{tiny}
\end{center}
\end{table}

\section{Results}
\label{results}

This section presents the results obtained with the GCM coupled with the haze parametrization. All figures and maps are shown using the new IAU convention, spin north system for definition of the North Pole \citep{Buie:97,Zang:15}, that is with spring-summer in the northern hemisphere during the 21th Century. Here we focus on model predictions in July 2015. We first compare the two reference simulations, then we show the corresponding ranges of UV and VIS opacities and we perform sensibility studies.

\subsection{Reference simulation 1: No South Pole N$_{2}$ condensation}


The predictions of the state of the atmosphere in July 2015 remain unchanged compared to what is shown in \citet{Forg:16}, since haze particles are not radiatively active and since their sedimentation on Pluto's surface does not impact the surface albedo. These processes could be taken into account in future GCM versions. 

In July 2015, the modeled surface pressure is found to be around 1 Pa. The nitrogen reservoir in Sputnik Planum at mid northern latitudes is under significant insolation during the New Horizon flyby (the subsolar latitude in July 2015 is 51.55\char23N), as well as the mid and high northern CH$_{4}$ frosts which sublime and become an important source of atmospheric CH$_{4}$, as described by \citet{Forg:16}. 

\begin{figure}[h]
\begin{center}
\includegraphics[scale=0.3]{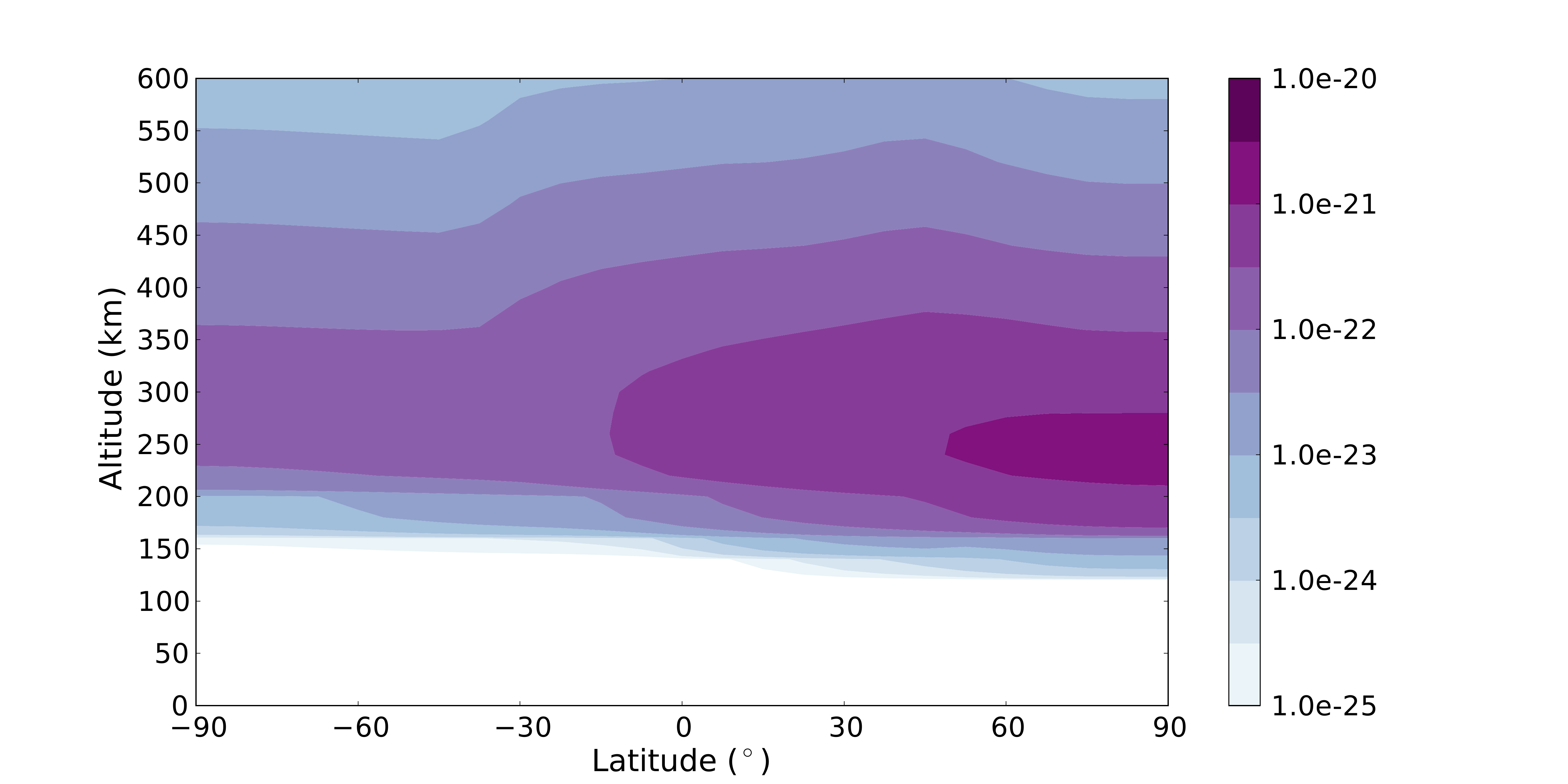}
\caption{Photolysis rate of CH$_{4}$ (g\,cm$^{-3}$\,s$^{-1}$) obtained with the reference simulation without South Pole N$_{2}$ condensation for July 2015 (color bar in log scale) \newline}
\label{photprecsection}
\end{center}
\end{figure}

According to equation \ref{reaction}, methane photolysis occurs at all latitudes but is more intense at locations where strong incoming flux of Lyman-$\alpha$ photons occurs, that is at high northern latitudes in July 2015. This is confirmed by \autoref{photprecsection}, showing the CH$_{4}$ photolysis rate as simulated in the GCM. All Lyman-$\alpha$ photons are absorbed above 150 km altitude. The maximum photolysis rate is is typically around 1.3$\times10^{-21}$ g\,cm$^{-3}$\,s$^{-1}$ and is obtained at 250 km altitude above the North Pole. 

\begin{figure}[h]
\begin{center}
\includegraphics[scale=0.2]{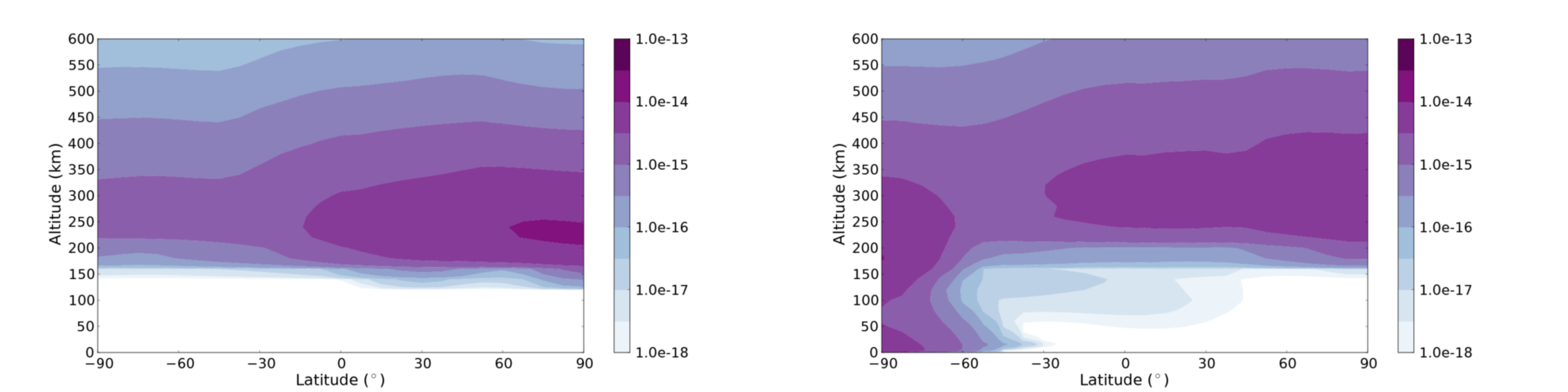}
\caption{Zonal mean latitudinal section of haze precursor density (g\,cm$^{-3}$) obtained with the reference simulation without (left) and with (right) South Pole N$_{2}$ condensation (color bar in log scale) \newline}
\label{prechazesection}
\end{center}
\end{figure}

Haze precursors formed by CH$_{4}$ photolysis are then transported by general circulation in the GCM. As shown by \citet{Forg:16}, the fact that N$_{2}$ ice is entirely sequestered in the Sputnik Planum basin and does not condense elsewhere leads to very low meridional wind velocities in the atmosphere and a weak meridional circulation.
Consequently, haze precursors are not transported fast towards the surface by circulation. In 2015, with a lifetime of 18 sols, the haze precursors are still confined to high altitudes above 140 km, and are in larger amount in northern latitudes where most of the photolysis of CH$_{4}$ occurs (\autoref{prechazesection}). 

\begin{figure}[h]
\begin{center}
\includegraphics[scale=0.2]{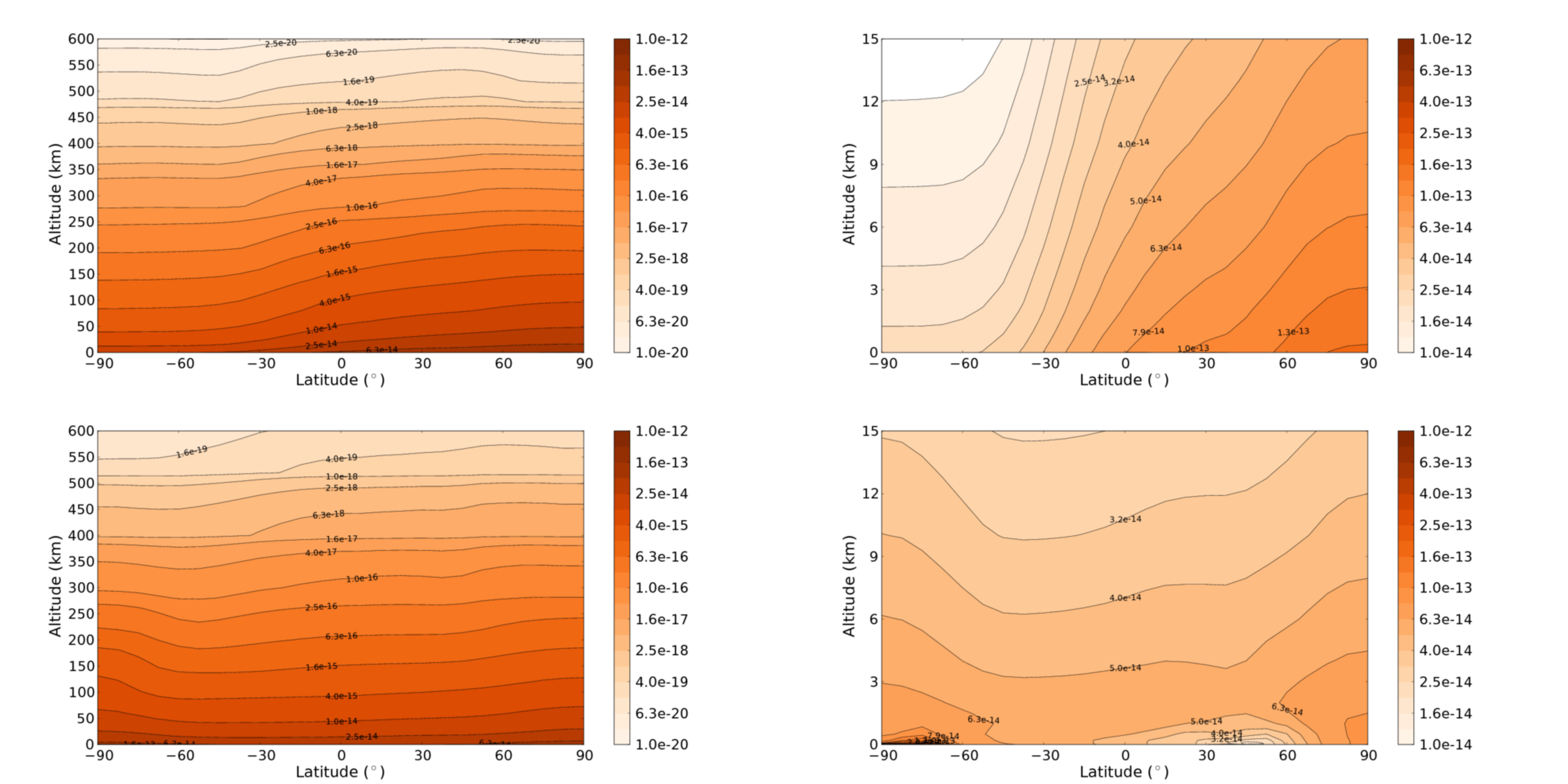}
\caption{Zonal mean latitudinal section of haze aerosol density (g\,cm$^{-3}$) obtained with the reference simulation for July 2015 without (top) and with (bottom) South Pole N$_{2}$ condensation (color bar in log scale). The right panels correspond to a zoom in the lowest 15 km above the surface.\newline}
\label{hazesection}
\end{center}
\end{figure}

\autoref{hazesection} shows the zonal mean latitudinal section of haze density predicted in July 2015. 
The aerosols formed above 150 km slowly fall towards the surface, and accumulate in the first kilometers above the surface, due to the decrease of sedimentation velocity with atmospheric pressure. 
The haze obtained extends at high altitudes. The density decreases with the altitude but remains non-negligible with values up to 4$\times10^{-19}$ g\,cm$^{-3}$ at 500 km altitude.
In this case, the meridional circulation is quite weak: the diurnal condensation and sublimation of N$_{2}$ ice in Sputnik region only impacts the circulation in the first km above the surface, and at higher altitudes, the circulation is forced by the radiative heating (the northern CH$_{4}$ warms the atmosphere, leading to a transport of this warm air from the summer to the winter hemisphere) inducing low meridional winds. Consequently, the general circulation does not impact the haze distribution, which is dominated by the incoming flux and the sedimentation velocity. In other words, the vertical and meridional atmospheric motions are not strong enough to signicantly push and impact the latitudinal distribution of the haze composed of 50 nm particles: the haze density in the atmosphere is always higher at the summer pole, where a stronger CH$_{4}$ photolysis occurs. 

In the summer hemisphere, the haze density is typically 2-4$\times10^{-15}$ g\,cm$^{-3}$ at 100 km altitude while it reaches 1-2$\times10^{-13}$ g\,cm$^{-3}$ above the surface.

\begin{figure}[h]
\begin{center}
\includegraphics[scale=0.35]{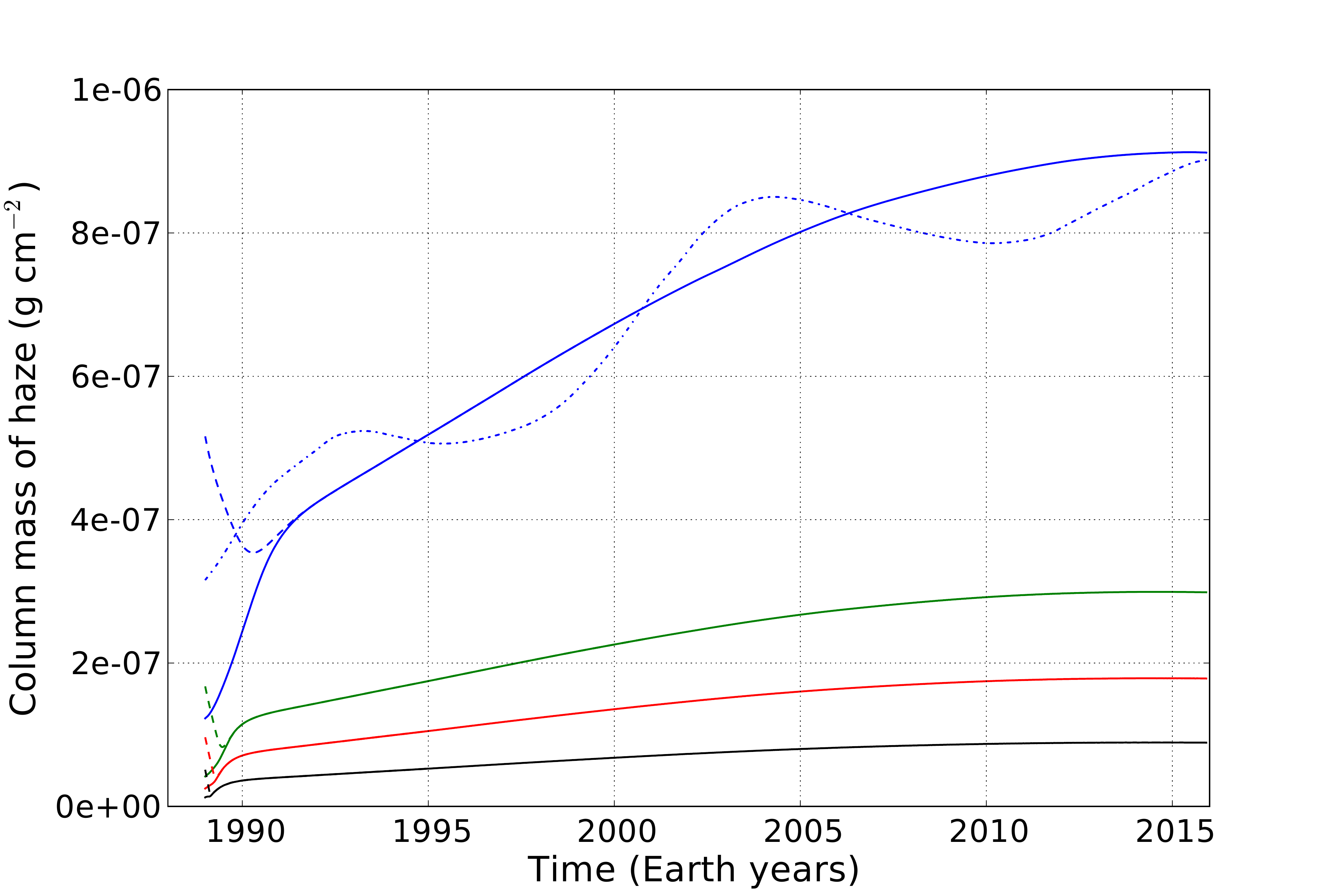}
\caption{Evolution of the mean column atmospheric mass of haze aerosols (g\,cm$^{-2}$) from 1988 to 2016 obtained with different particle radius in the reference simulation without South Pole N$_{2}$ condensation: 10 nm (blue), 30 nm (green), 50 nm (red) and 100 nm (black). The dashed lines correspond to similar simulations started with a higher initial amount of haze. With 50 nm particles (red curve), the mass of haze reaches an equilibrium within less than one year. The dash-dotted line corresponds to the 10 nm case with the real variable initial Lyman-$\alpha$ flux (at Earth).\newline}
\label{evolhaze}
\end{center}
\end{figure}

\autoref{evolhaze} shows the evolution of the mean column atmospheric mass of haze aerosols since 1988. Assuming a constant initial flux of Lyman-$\alpha$ (at Earth) and a particle radius of 50 nm, the column mass of haze reaches a peak of 1.8$\times10^{-7}$ g\,cm$^{-2}$ in 2015. 
Because the transport of haze is dominated by its sedimentation, the column mass of haze directly depends on the sedimentation velocity of the haze particles. As shown by equation \ref{stokes}, the sedimentation velocity decreases when pressure increases, hence the increase of column mass of haze, in line with the threefold increase of surface pressure since 1988. Note that this trend still applies when considering the real and variable initial Lyman-$\alpha$ flux at Earth between 1988 and 2015, as shown by \autoref{evolhaze}.

\begin{figure}[h]
\begin{center}
\includegraphics[scale=0.2]{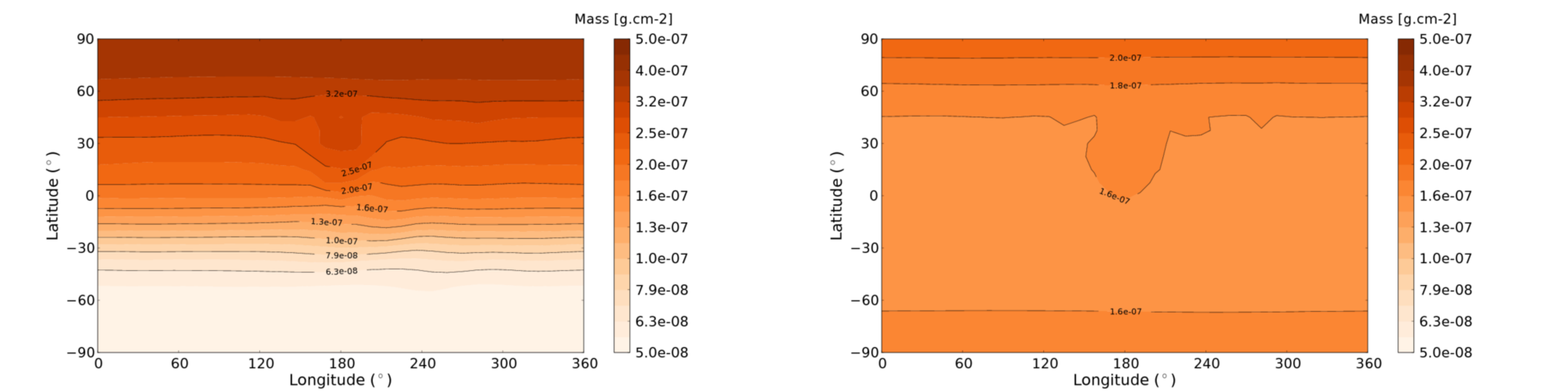}
\caption{Column atmospheric mass map of haze aerosols (g\,cm$^{-2}$) obtained with the reference simulation without (left) and with (right) South Pole N$_{2}$ condensation \newline}
\label{massmaphaze}
\end{center}
\end{figure}

\autoref{massmaphaze} shows the column atmospheric mass of haze aerosols. In line with the previous results, the column mass obtained is higher at the North Pole than at the South Pole by one order of magnitude, due to the maximum haze production in the summer hemisphere. The column mass of haze reaches 3.9$\times10^{-7}$ g\,cm$^{-2}$ at the North Pole.


\subsubsection{Reference simulation 2: with South Pole N$_{2}$ condensation}
\label{sec:polarcap}

The sublimation of N$_{2}$ in mid northern latitudes (Sputnik region and the latitudinal band) and its condensation in the winter hemisphere induce an atmospheric flow from the northern to the southern hemisphere, and thus a stronger meridional circulation than in the reference simulation without South Pole N$_{2}$ condensation, although the latitudinal winds remain relatively weak \citep{Forg:16}. 
Although the atmospheric methane is more mixed in the atmosphere in this case, the state of the atmosphere remains similar to the reference simulation without South Pole N$_{2}$ condensation. The surface pressure is increasing before 2015 and reaches 1 Pa in 2015. 

Because of the condensation flow from the northern to the southern hemisphere, the air in the upper atmosphere is transported along with the haze precursors from the summer atmosphere to the winter atmosphere.
As shown on \autoref{prechazesection}, the characteristic decay time of haze precursors (18 sols) is sufficient for some of the precursors to be transported from the summer to the winter hemisphere where the descending branch bring them at lower altitudes down to the surface.




As a consequence of that, more haze is formed in the winter hemisphere than in the reference simulation without N$_{2}$ condensation flow, which compensates the haze production in the summer hemisphere due to the higher CH$_{4}$ photolysis rate. It leads to a similar haze density at all latitudes, as shown by \autoref{hazesection}. 
The haze density is typically 4$\times10^{-15}$ g\,cm$^{-3}$ at an altitude of 100 km, which is similar to the reference simulation without the condensation flow. The haze remains latitudinally well dispersed down to 3 km, where the meridional circulation driven by the N$_{2}$ condensation flow affects the haze distribution: the haze is pushed towards southern latitudes by the N$_{2}$ ice sublimation above the N$_{2}$ frost latitudinal band and Sputnik Planum, avoiding an accumulation of haze at the mid and high northern latitudes. Between -70\char23S and -90\char23S, haze particles in the first layers are suctioned towards the surface of the N$_{2}$ polar cap. 
The haze reaches a density of about 5-20$\times10^{-12}$ g\,cm$^{-3}$ below 1 km in the winter hemisphere, and 3-6$\times10^{-14}$ g\,cm$^{-3}$ in the summer hemisphere, which is twice less compared to the reference simulation without the condensation flow.

In line with the previous results, the column mass of haze in the simulation with condensation flow shown on \autoref{massmaphaze} (right figure) is well dispersed on Pluto, with small variations: in the summer atmosphere, the mass is about 2$\times10^{-7}$ g\,cm$^{-2}$, but it is slightly less at low and mid latitudes because the haze above the surface is transported towards the south polar cap, and slightly more at the North Pole because the haze is not impacted by the N$_{2}$ ice sublimation and transport which occur at lower latitudes.

As in the previous simulation without South Pole N$_{2}$ condensation, the mean column mass of haze increases with surface pressure. In 2015, a similar averaged column mass of haze is obtained. Slight discrepancies are found due to slightly different surface pressures to first order \citep{Forg:16}, and to the different circulation to second order. 


\subsection{Haze opacity}
\label{opacity}

In order to better quantify the amount of haze formed on Pluto and compare with the observations as well as with the Titan and Triton cases, one can compute the total column opacity and the line of sight opacity of the haze (as a diagnostic of the results). Here we focus on the opacity at UV ($\lambda$ = 150 nm) and visible ($\lambda$ = 550 nm) wavelengths for sake of comparison with the data recorded by the UV spectrometer Alice and the Ralph and LORRI instruments on board New Horizons. Assuming a homogeneous size and extinction efficiency for the aerosols in Pluto's atmosphere, the opacity $\tau_{\lambda}$ for a given wavelength $\lambda$ is directly proportional to the atmospheric column mass of aerosols:

\begin{equation}
\tau_{\lambda}=\alpha.M \quad \quad with \quad \quad
\alpha=\frac{3}{4}\frac{Q_{ext,\lambda}}{\rho_{aer}r_{eff}}
\label{tau}
\end{equation}

where $Q_{ext}$ is the aerosol extinction efficiency, $r_{\mbox{eff}}$ the aerosol particle effective radius, $\rho_{aer}$ the aerosol density and $M$ is the atmospheric column mass of aerosol in kg\,m$^{-2}$. 

\subsubsection{Spherical particles}

Assuming that the haze on Pluto is composed of spherical particles and behaves like the detached haze layer on Titan, we used a Mie code to generate single scattering extinction properties for different spherical particle sizes. The code takes into account a modified gamma size distribution of particles with the considered effective radius and an effective variance $\nu_{eff}$ = 0.3, as well as the optical indices of \citet{Rann:10}. These indices have been updated from \citet{Khar:84} thanks to new sets of Cassini observations. For 50 nm particles, we obtain an extinction efficiency Q$_{ext}$ of 2.29 in UV and 0.19 in visible wavelengths. Using equation \ref{tau} with a density of aerosol material of 800 kg\,m$^{-3}$, we find that the haze column opacity in July 2015 reaches 0.077-0.17 (UV) and 0.0064-0.014 (VIS) in the summer hemisphere, in the reference simulation without South Pole N$_{2}$ condensation. In the simulation with South Pole N$_{2}$ condensation, the opacities are 0.064-0.086 (UV) and 0.0053-0.0071 (VIS) in the summer hemisphere. 

\subsubsection{Fractal particles}

The case of fractal particles can also be discussed. On Titan, an upper limit of the maximum equivalent mass sphere radius (or bulk radius) of fractal particles in the detached haze layer has been estimated to 300 nm, containing up to 300 monomers \citep{Lars:14}, while larger particles containing a higher number of monomers are mostly found in the main haze atmosphere of Titan, at lower altitudes. 
In fact, some aerosols of the detached haze layer on Titan are large aggregates that grow within the main haze layer at lower altitudes and that are lift up back to the detached layer by ascending currents occurring in the summer hemisphere \citep{Rann:02,Lebo:09}. On Pluto, such mechanisms are not likely to occur because of the thin atmosphere, and the size of fractal particles, if formed, should be limited.
Consequently, we consider only a small fractal particle with a limited amount of monomers.

Fractal particles have a different optical behavior compared to spherical particles. As shown by the figure 10 in \citet{Lars:14}, the optical depth of a 1 $\mu$m fractal particle is strongly dependent on the considered wavelength and decreases from the UV to the near infrared, while the optical depth of a similar sized spherical particle remains quite constant with the wavelength.
One can use equation \ref{tau} to calculate the opacity of fractal particles with $Q_{ext}$ the aerosol extinction efficiency (referred to the equivalent mass sphere), $r_{\mbox{eff}}$ the equivalent mass sphere radius of the particle and $\rho_{aer}$ the density of the material (or density of the monomers).
Here we used a mean field model of scattering by fractal aggregates of identical spheres \citep{Bote:97,Rann:97} to estimate the extinction efficiency of fractal particles. 
From the number of monomers N and the monomers radius $r_{m}$, on can calculate the equivalent mass sphere radius of the corresponding fractal particle, given by $R_{s}$ = N$^{\frac{1}{3}} \times r_{m}$. Using these parameters and the fractal dimension of the particle, the model computes $Q_{ext}$ by dividing the extinction cross section of the particle by the geometrical cross section of the equivalent mass sphere ($\pi$\,$R_{s}^{2}$). 

Here we compare the opacities obtained in the reference simulations when considering spherical or fractal particles. We consider fractal particles composed of 50 nm monomers, with a fractal dimension equal to 2 and with a bulk radius of 100 nm and 232 nm (N=8 and N=100 monomers respectively).
The model gives an extinction efficiency Q$_{ext}$ of 4.1 in the UV and 0.49 in the visible wavelengths for the 100 nm fractal particle and 7.2 in the UV and 1.93 in the visible wavelengths for the 232 nm fractal particle. The resulting nadir opacities are summarized in \autoref{tabop_r} and limb opacities are shown on \autoref{opacity_limb_ref}. 
The opacities obtained for fractal particles are higher than for spherical particles in the visible, with a factor of 1.3 for the 100 nm and 2.2 for the 232 nm particle but lower in the UV with a factor of 0.9 and 0.7 respectively for the 100 nm and the 232 nm particle. This is shown by \autoref{opacity_limb_ref}.

As shown in \autoref{tabop_r}, the visible nadir opacity obtained in the summer hemisphere are in the range of what is estimated from New Horizons observations (0.004-0.012, \citet{Ster:15,Glad:16}) in both the spherical and the 100 nm fractal cases, and in both reference simulations. Values of the 232 nm fractal case are outside the observational range. The case of fractal particles composed of 10 nm particles is discussed in Section \ref{sec:sensib_r}.
 

\subsubsection{Line of sight opacity profiles}

\begin{figure}[h]
\begin{center}
\includegraphics[scale=0.2]{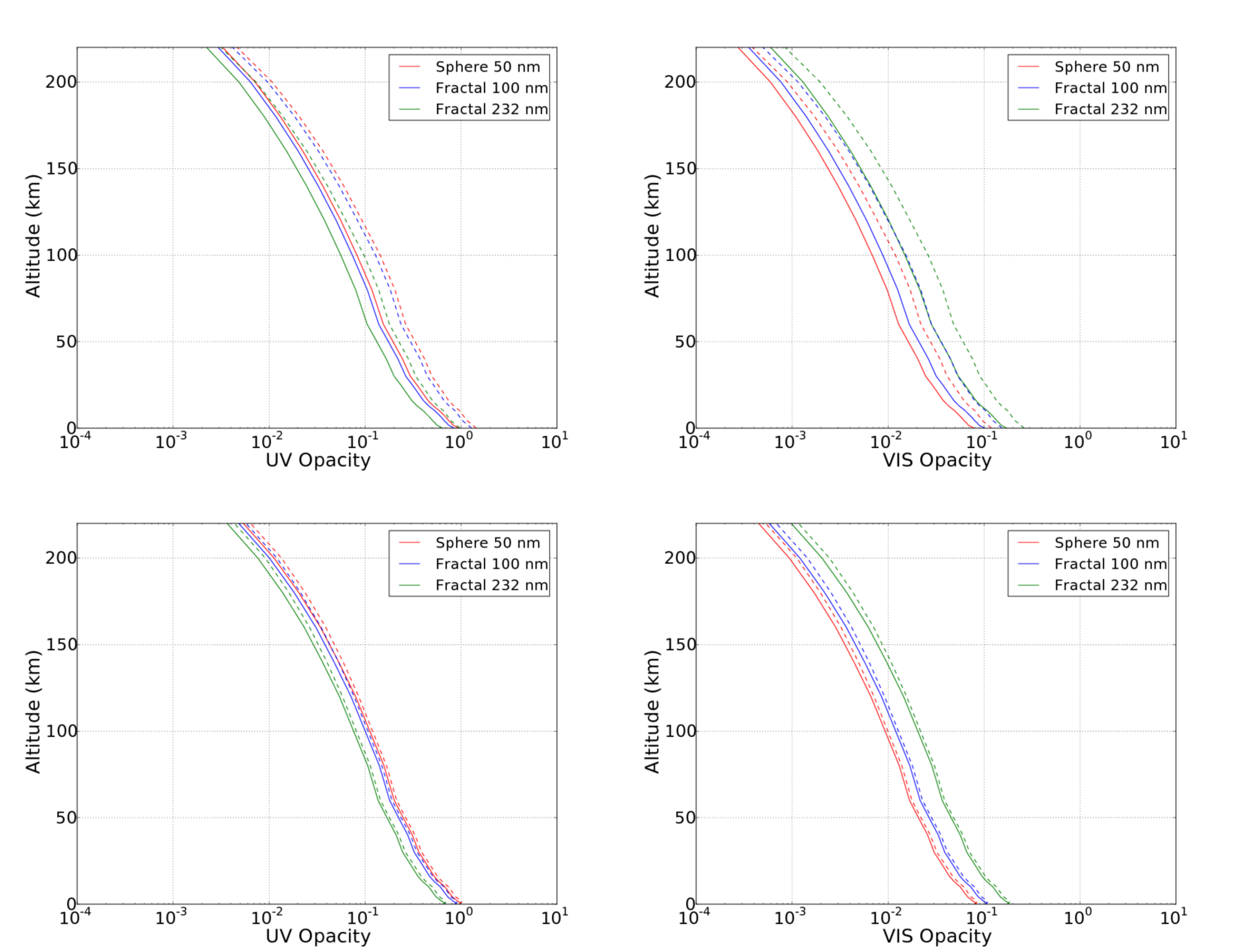}
\caption{Line of sight opacity profiles obtained with the GCM for the spherical and fractal cases, at the ingress (-163\char23E, 17\char23S, solid lines) and egress point (16\char23E, 15\char23N, dashed lines) of Pluto's solar occultation, for the reference simulation without (top) and with (bottom) South Pole N$_{2}$ condensation. Left and right are the results in UV and VIS wavelength respectively. The red curve is the reference simulation with 50 nm spherical particles. The blue and green curves correspond to the fractal cases with R$_{s}$=100 nm / N=8 and R$_{s}$=232 nm / N=100 respectively. \newline}
\label{opacity_limb_ref}
\end{center}
\end{figure}

\autoref{opacity_limb_ref} shows the line of sight opacity profiles in the UV and in the visible wavelengths obtained for both reference simulations at the ingress and the egress points of Pluto's solar occulation by New Horizons. The profiles are computed using an onion peeling method and considering that the line of sight only crosses one GCM atmospheric column. 

Generally speaking, few differences are obtained between both reference simulations. The difference of opacity between the egress point (which is above the equator at the latitude 15\char23N) and the ingress point (which is below the equator at the latitude 17\char23S) are larger for the simulation without South Pole N$_{2}$ condensation, because of the higher haze density in the summer hemisphere shown in \autoref{hazesection}. 



\newpage

\subsection{Sensitivity studies}
\label{sec:sensib}

The poor constraint on haze properties on Pluto gives us a flexibility to explore further other scenarios for Pluto's haze. In this section, the haze parametrization is tested with different precursor lifetimes and sedimentation radius. We also discuss the possible values for K$_{CH4}$ in the parametrization.
One objective is to investigate if another set of haze parameters can cause a more realistic aerosol distribution and concentration in the sunlit equatorial and summer atmosphere, compared to the observations. 
In addition, the sensitivity study aims to “bracket” the reality of Pluto's haze by analyzing extreme cases and compare them to both reference simulations.
First, it has been checked that the haze production is insensitive to the amount of CH4 present in the upper atmosphere. Although the amount of CH4 molecules decreases in the upper atmosphere due to the absorption of incident photons and photolysis reactions, this loss remains negligible compared to the total amount of CH4 in Pluto's atmosphere. In addition, the production of haze precursors still occurs at high altitudes above 100 km even for low values of CH$_{4}$ mixing ratio. The ratio between the production rate of precursors at 100 km and the rate at 220 km (top of the model) becomes higher than 1\% for a mean CH$_{4}$ mixing ratio of 0.04\%, which is one order of magnitude less than the typical values found on Pluto. This confirms that the reaction is photon-limited and that different (and realistic) CH4 mixing ratio will not impact haze production and distribution. 

\subsubsection{Sensitivity to characteristic time for aerosol growth}
\label{sec:sensib_tau}

\begin{figure}[h]
\begin{center}
\includegraphics[scale=0.3]{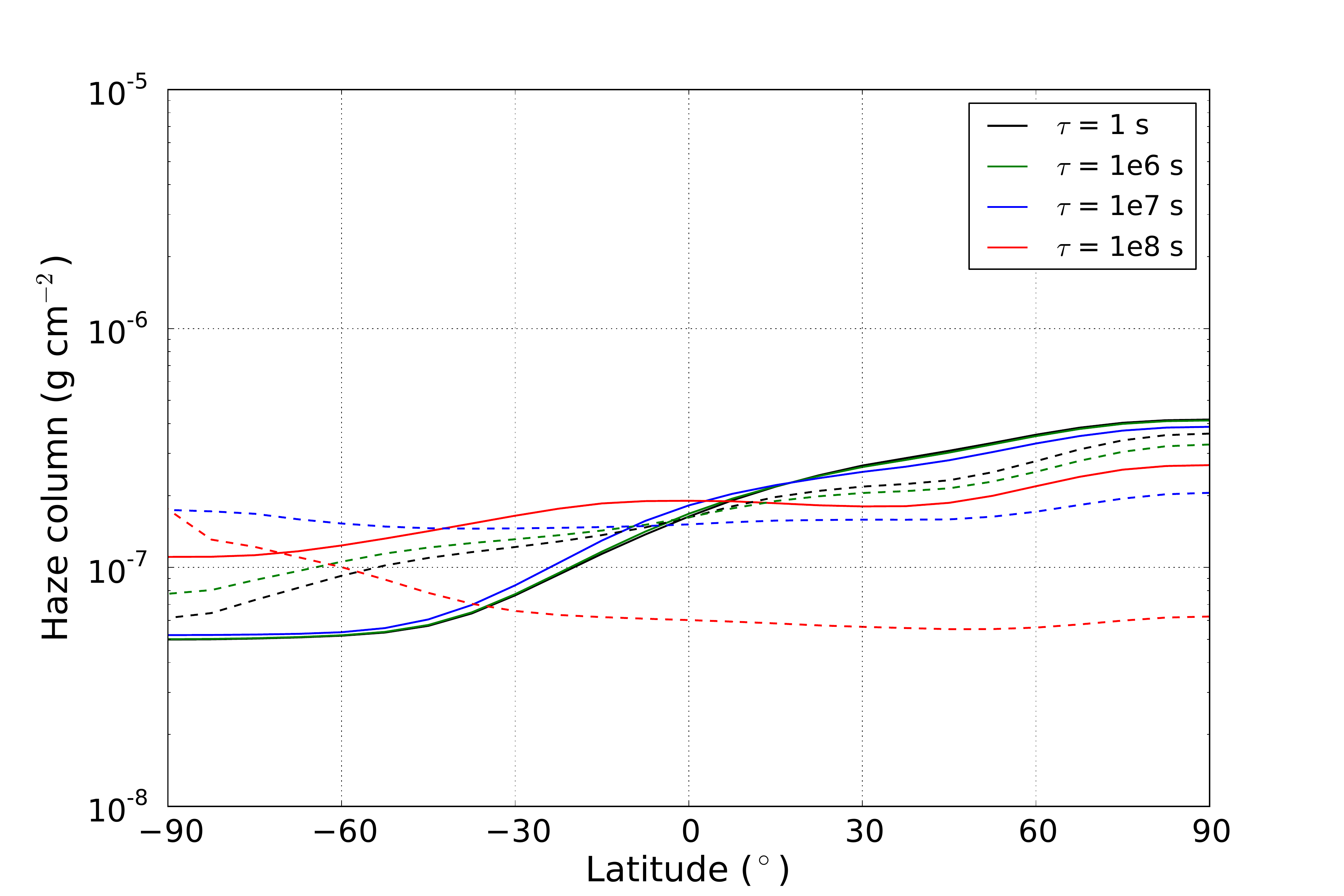}
\caption{Zonal mean of column atmospheric mass of haze aerosols (kg\,m$^{-2}$) obtained for July 2015 with different times for aerosol growth  $\tau$ (s), for the simulations without (solid lines) and with (dashed lines) South Pole N$_{2}$ condensation. \newline}
\label{masshaze_tau}
\end{center}
\end{figure}

\begin{figure}[h]
\begin{center}
\includegraphics[scale=0.2]{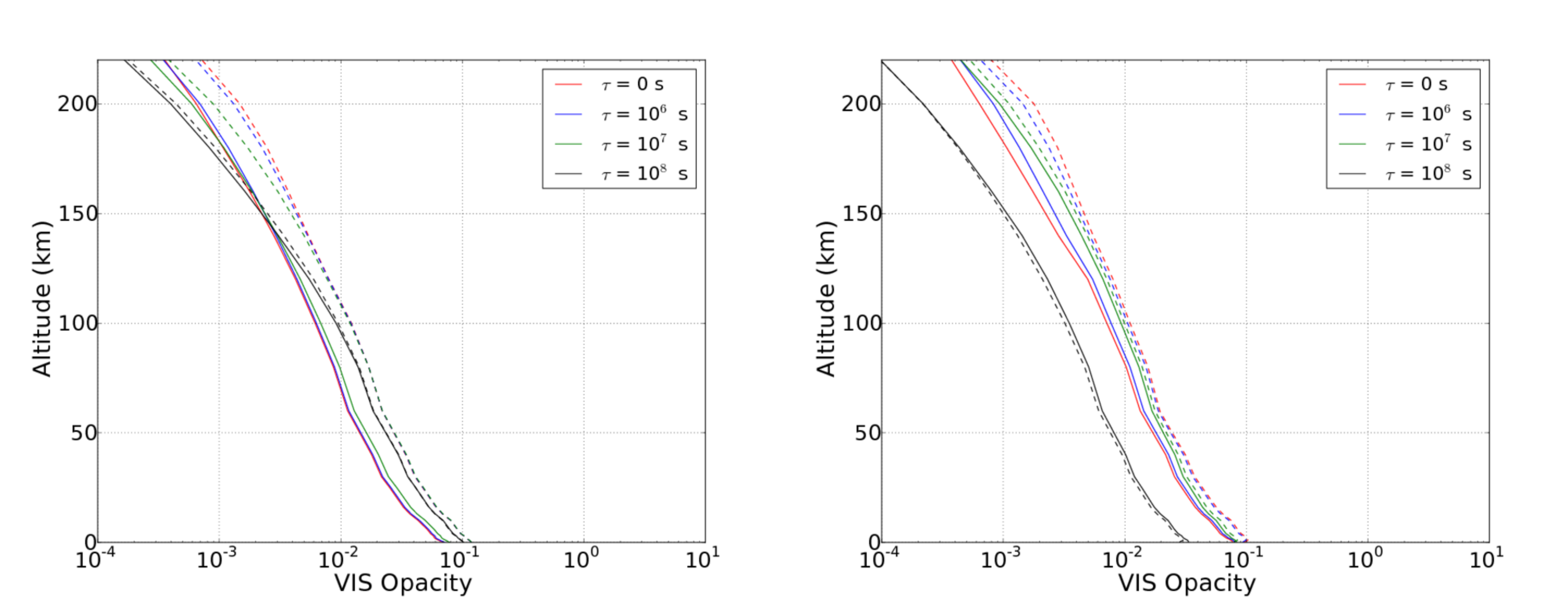}
\caption{Line of sight opacity profiles in VIS wavelength obtained with the GCM with different times for aerosol growth, at the ingress (-163\char23E, 17\char23S, solid lines) and egress point (16\char23E, 15\char23N, dashed lines) of Pluto's solar occultation, for the simulations without (left) and with (right) South Pole N$_{2}$ condensation \newline}
\label{opacity_limb_tau}
\end{center}
\end{figure}

The characteristic time for aerosol growth, defined in Section \ref{conversion}, is challenging to estimate. Here we consider two possible extreme values in the model. If this time is set to 1 second, this means that precursors are instantaneously converted into haze aerosols in the upper atmosphere where CH$_{4}$ photolysis occurs. This remains acceptable since photolysis and photochemistry can actually occur at much higher altitudes above the model top. An upper value up to several terrestrial years seems reasonable considering the number of years simulated and will allow precursors to be more mixed in the entire atmosphere.
Here we compare simulation results obtained with different characteristic times for aerosol growth (\autoref{masshaze_tau} and \autoref{opacity_limb_tau}): 1 s (haze directly formed from photolysis reactions), 10$^{6}$ s (1.81 Pluto sols), 10$^{7}$ s (18.12 sols, reference simulations), 10$^{8}$ s (181.20 sols, that is about 3 terrestrials years). The rest of the settings remain similar to both reference simulations. 

In the simulations without South Pole N$_{2}$ condensation, using 1-10$^{7}$ s leads to similar column mass of haze, as shown by \autoref{masshaze_tau}. With a lifetime of 10$^{8}$ s, the precursors have enough time to be transported by the circulation induced by radiative heating from the summer to the winter hemisphere, and at lower altitudes. It results in a better dispersed haze density at all latitudes, a lower mass in the summer hemisphere, and thus similar egress and ingress line of sight opacities, as shown on \autoref{opacity_limb_tau}.

In the simulations with South Pole N$_{2}$ condensation, the longer the precursor lifetime, the more they are transported by radiative heating towards the 
winter hemisphere and by the descending circulation branch towards the surface of the winter polar cap. Thus, the haze tends to accumulate in the winter hemisphere and in lower amounts if long lifetimes are considered, and in the summer hemisphere in larger amounts otherwise. 

The difference of opacity obtained between the egress and the ingress points is larger for low lifetimes and conversely, as shown on \autoref{opacity_limb_tau}.

\subsubsection{Sensitivity to particle radius}
\label{sec:sensib_r}

\begin{figure}[h]
\begin{center}
\includegraphics[scale=0.3]{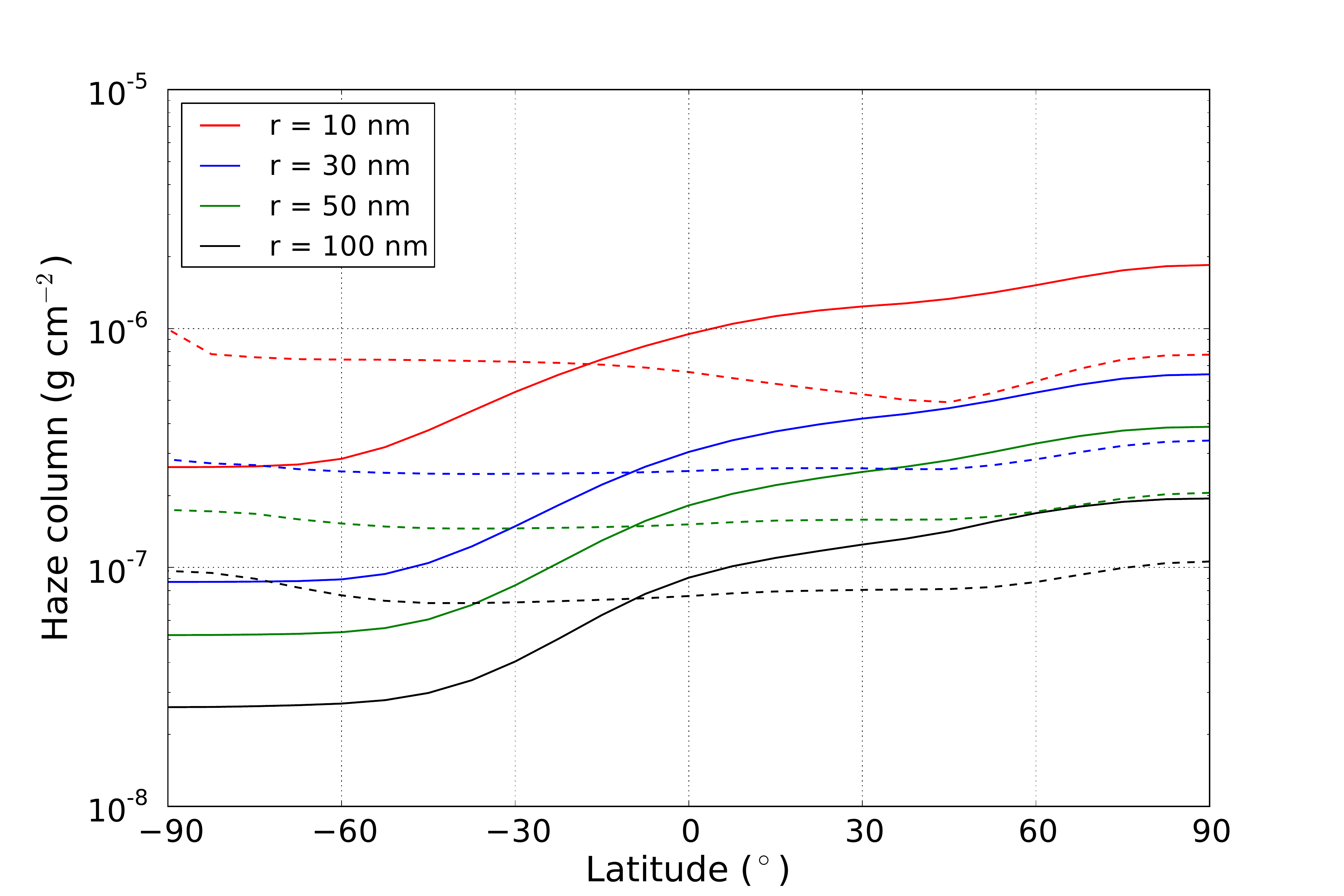}
\caption{Zonal mean of column atmospheric mass of haze aerosols (kg\,m$^{-2}$, log scale) obtained with different particle radii, for the simulations without (solid lines) and with (dashed lines) South Pole N$_{2}$ condensation.}
\label{masshaze_r}
\end{center}
\end{figure}

\begin{figure}[h]
\begin{center}
\includegraphics[scale=0.2]{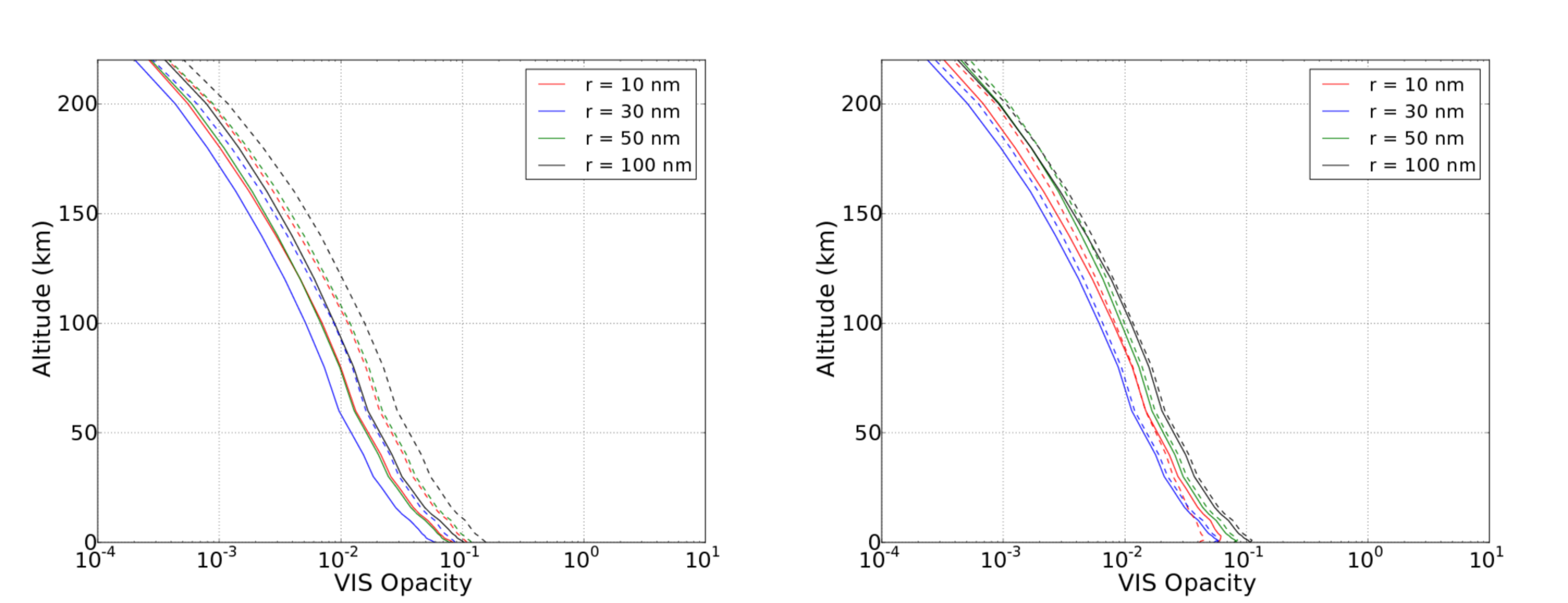}
\caption{Line of sight opacity profiles in VIS wavelength obtained with the GCM for different spherical particle radii, at the ingress (-163\char23E, 17\char23S, solid lines) and egress point (16\char23E, 15\char23N, dashed lines) of Pluto's solar occultation, for the simulations without (left) and with (right) South Pole N$_{2}$ condensation \newline}
\label{opacity_limb_r}
\end{center}
\end{figure}

\begin{figure}[h]
\begin{center}
\includegraphics[scale=0.2]{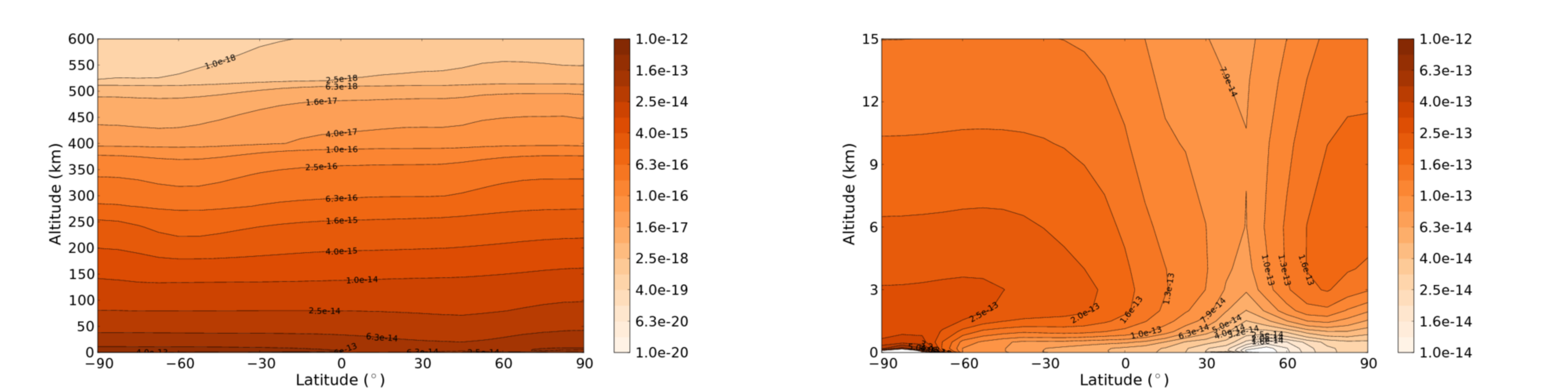}
\caption{Zonal mean latitudinal section of haze aerosol density (g\,cm$^{-3}$) obtained with the simulation for July 2015 with condensation flow and a particle radius of 10 nm (color bar in log scale). The right panel correspond to a zoom in the lowest 15 km above the surface.\newline}
\label{hazesection_r10_cap}
\label{lastfig}
\end{center}
\end{figure}

The uniform and constant radius of aerosol particles is a parameter that strongly controls the aerosol sedimentation and opacity in the GCM. As shown by equation \ref{stokes} in Section \ref{hazeproperties}, a smaller particle radius induce a lower haze sedimentation velocities and thus a higher mass of haze in the atmosphere.
Here we compare eight simulations: the reference simulations (50 nm particles, with and without condensation flow) and simulations performed with particle sizes of 10, 30 and 100 nm (with and without condensation flow). We compare the column atmospheric mass obtained (\autoref{masshaze_r}), the limb opacities (\autoref{opacity_limb_r}) and the nadir opacities (\autoref{tabop_r}). These simulations correspond to the four first lines of \autoref{tabop_r}. The six last lines of \autoref{tabop_r} show the nadir opacities obtained from the simulations with 10 nm and 50 nm particles,   but considering fractal particles (four cases with 10 nm monomers and two cases with 50 nm monomers). Haze aerosol density is also shown for the simulation with condensation flow and with a particle radius of 10 nm (\autoref{hazesection_r10_cap}). 

Aerosol particles with radii of 10, 30, 50 and 100 nm typically fall from 200~km down to the surface in 1110, 370, 220 and 111 Earth days respectively. Basically, this corresponds to the time needed to reach an equilibrated mass of haze in the atmosphere.
As shown by \autoref{masshaze_r}, the latitudinal mass distribution is not impacted by the considered size of the particle. The column mass of haze is driven by the sedimentation velocity and the mass ratios correspond to the particle size ratios. This is also shown by \autoref{evolhaze}.

As shown by \autoref{tabop_r} and \autoref{opacity_limb_r}, the nadir and limb opacities remain in the same order of magnitude for the simulations performed with different particle radii. Lower opacities are obtained with a particle radius of 30 nm. We also investigated nadir opacities for fractal particles with a bulk radius of 22, 46, 100 and 200 nm, respectively composed of 10, 100, 1000 and 8000 monomers of 10 nm radius. As discussed in Section \ref{background}, the 200 nm fractal particle is the best hypothesis for the particle shape and size in order to fit the observations. Here we find that the nadir visible opacities obtained in this case are higher than the upper observational limit (see \autoref{tabop_r}). Realistic values are obtained for the other smaller particles.

\autoref{opacity_limb_r} show the line of sight visible opacities obtained for different spherical particle radii. Generally speaking, the profiles have similar shapes because changing the particle radius does not affect the haze distribution but only the mass of haze in the atmosphere, due to the change of sedimentation velocity. 
However, for 10 nm particles, the opacities at ingress are significantly higher than at egress below 50 km, which is not the case for higher radii. 
This is because the particles are lighter and have more time to be transported by the circulation towards the winter hemisphere before sedimentation to the surface. Thus, the change of haze distribution due to the condensation flow below 50 km altitude is more pronounced for this 10 nm case. This is highlighted by \autoref{hazesection_r10_cap} which shows the 10 nm haze particles density in the simulation with condensation flow. In the first kilometers above the surface, a peak of density is obtained at the South Pole. In addition, above 2 km altitude, the haze also accumulates at the North Pole, pushed away by the condensation flow. 


\begin{table}[h]
\begin{center}
\begin{tiny}
\begin{tabular}{m{1.5cm}m{0.5cm}m{0.8cm}m{0.8cm}m{1.5cm}m{1cm}m{1cm}m{1.5cm}m{1cm}m{1cm}}
\hline
   & & & & \multicolumn{3}{c}{Without winter polar cap} & \multicolumn{3}{c}{With winter polar cap} \\ 
\hline
Radius & N$_{m}$ & Q$_{ext}$ UV & Q$_{ext}$ VIS & Aerosol mass (g\,cm$^{-2}$) & UV opacity & VIS opacity & Aerosol mass (g\,cm$^{-2}$) & UV opacity & VIS opacity \tabularnewline
r = 10 nm & 1 & 0.35 & 0.007 & $9.5-18\times10^{-7}$ & 0.31-0.59 & 0.0062-0.012 & $4.9-7.8\times10^{-7}$ & 0.16-0.26 & 0.0032-0.0051 \\
r = 30 nm & 1 & 1.54 & 0.05 & $3.0-6.5\times10^{-7}$ & 0.14-0.31 & 0.0047-0.010 & $2.5-3.4\times10^{-7}$ & 0.12-0.17 & 0.0039-0.0053 \\
r = 50 nm (reference) & 1 & 2.29 & 0.19 & $1.8-3.9\times10^{-7}$ & 0.077-0.17 & 0.0064-0.014 & $1.5-2.0\times10^{-7}$ & 0.064-0.086 & 0.0053-0.0071 \\
r = 100 nm & 1 & 2.67 & 1.01 & $0.9-1.9\times10^{-7}$ & 0.023-0.048 & 0.0085-0.018 & $0.75-1.1\times10^{-7}$ & 0.019-0.028 & 0.0071-0.010 \\
R$_{s}$ = 22 nm r~=~10~nm& 10 & 0.84 & 0.018 & $9.5-18\times10^{-7}$ & 0.34-0.64 & 0.0073-0.014 & $4.9-7.8\times10^{-7}$ & 0.18-0.28 & 0.0038-0.0060 \\
R$_{s}$ = 46 nm r~=~10~nm& 100 & 2.06 & 0.052 & $9.5-18\times10^{-7}$ & 0.40-0.76 & 0.010-0.019 & $4.9-7.8\times10^{-7}$ & 0.21-0.33 & 0.0052-0.0083 \\
R$_{s}$ = 100 nm r~=~10~nm& 1000 & 4.65 & 0.15 & $9.5-18\times10^{-7}$ & 0.41-0.78 & 0.013-0.025 & $4.9-7.8\times10^{-7}$ & 0.21-0.34 & 0.0069-0.0110 \\
R$_{s}$ = 200 nm r~=~10~nm& 8000 & 9.44 & 0.38 & $9.5-18\times10^{-7}$ & 0.42-0.80 & 0.017-0.032 & $4.9-7.8\times10^{-7}$ & 0.22-0.35 & 0.0087-0.0139 \\
R$_{s}$ = 100 nm r~=~50~nm& 8 & 4.10 & 0.49 & $1.8-3.9\times10^{-7}$ & 0.069-0.15 & 0.0083-0.018 & $1.5-2.0\times10^{-7}$ & 0.058-0.077 & 0.0069-0.0092 \\
R$_{s}$ = 232 nm r~=~50~nm& 100 & 7.20 & 1.93 & $1.8-3.9\times10^{-7}$ & 0.052-0.11 & 0.014-0.030 & $1.5-2.0\times10^{-7}$ & 0.044-0.058 & 0.0117-0.0156 \\
\hline 
\end{tabular}
\caption{Haze aerosol opacities obtained at nadir in the summer hemisphere in the GCM, for four particle radii and for both climate scenarios with and without South Pole N$_{2}$ condensation. The time for aerosol growth used is $10^{7}$ s. The particles with a number of monomers N$_{m}$ equal to 1 are spherical particles, otherwise they are fractal particles (R$_{s}$ is the bulk radius, $r$ is the monomer radius). The first four fractal particles are composed of 10 nm monomers, and the last two are composed of 50 nm monomers. \newline}
\label{tabop_r}
\label{lasttable}
\end{tiny}
\end{center}
\end{table}

\newpage

\subsubsection{Sensitivity to the mass of aerosols}
\label{sec:sensib_mass}

The haze production rate used in the reference simulations corresponds to an optimal scenario where the photolysis of one molecule of CH$_{4}$ gives one carbon atom available for the production of haze ($K_{CH4}$=1). However, the carbon atoms collected from CH$_{4}$ photolysis may form different gaseous species and slow down tholins production. As an example, \citet{McKa:01} suggest that the tholins production is about 25 less than the photolysis rate of methane. Therefore, lower values of $K_{CH4}$ remain possible and would lead to a decrease of aerosol mass and thus of opacity. 


\section{Summary}
The parametrization of haze aerosols in the Pluto GCM consists of several steps: the photolysis of methane by the solar and IPM flux, the creation of haze precursors and their transport in the atmosphere, the conversion of precursors to haze aerosols and the sedimentation of the aerosols. 
The haze parametrization has been tested with 50 nm particles, a time for aerosol growth of 10$^{7}$ s, and for the two climate scenarios described in \citet{Forg:16}: with and without South Pole N$_{2}$ condensation (reference simulations). The sensitivity of the model to other particle sizes and times for aerosol growth has been explored.
Results show that the CH$_{4}$ photolysis occurs at all latitudes, with a maximum rate at high northern latitudes and around 250 km in altitude. In all simulations, the haze extends to high altitudes, comparable to what has been observed by New Horizons. From 200 km altitude upwards, the density decreases with the altitude by one order of magnitude every 100 km, leading to a density scale height of typically 40 km above 60 km altitude. This is comparable to the typical haze brightness scale height of 50 km observed by New Horizons \citep{Glad:16}.
Without South Pole N$_{2}$ condensation, the meridional atmospheric circulation is dominated by the radiative heating but remains weak, even in the first kilometers above the surface. The haze precursors remains at high altitudes and in larger amount  at high northern latitudes. This leads to a higher density of haze in the summer hemisphere, decreasing with the latitudes. 
With South Pole N$_{2}$ condensation, the circulation is also weak in the upper atmosphere, except above the South Pole where a descending branch of air driven by the condensation of N$_{2}$ transports the precursors to lower altitudes. This leads to a distribution of haze latitudinally more homogeneous with a slight peak of haze density above the South Pole. This peak is reiforced by the circulation in the first kilometers above the surface, which is more intense and able to move light aerosols from the northern hemisphere towards the South Pole. 
In both climate scenarios, because of the generally weak meridional circulation, the computed mean atmospheric column mass of haze remains similar, and primarily depends on the sedimentation velocity and thus on the pressure and the considered monomer radius. In our simulations, the initial flux of Lyman-$\alpha$ at Earth remains constant between 1990 and 2015, but even if we consider the variable initial flux of Lyman-$\alpha$, the flux of Lyman-$\alpha$ at Pluto remains relatively constant. Consequently, the mean column mass of haze follows the trend in surface pressure, that is an increase by a factor of 3 between 1990 and 2015. Haze particles with a small radius remain longer in the atmosphere before reaching the surface. In our simulations, the sedimentation fall of 10 nm particles lasts about 3 terrestrial years, which could be enough time to form fractal aggregates. 
The mean column atmospheric mass of haze on Pluto is difficult to assess because it depends on many parameters. First, it is depending on the photolysis rate and the complex recombinations of carbon and nitrogen atoms. The parametrization uses K$_{CH4}$ and K$_{N}$ equal to 1 and 1.5 to take these mechanisms into account. However, the production could be overestimated. In fact, New Horizons detected the presence of C$_{2}$H$_{2}$, C$_{2}$H$_{4}$ and maybe other carbon-based gas in Pluto atmosphere, which suggests another pathway for carbon atoms formed by CH$_{4}$ photolysis. In addition, HCN has been detected, and the irreversible nature of its formation may lead to less nitrogen atoms available for the haze formation.  
The column mass of haze also strongly depends on the sedimentation radius of the haze particle, and to a lesser extent on the lifetime of the haze precursors.
However, we computed the UV and VIS opacities of the haze as a diagnostic of our simulation results and in all simulation cases, the column visible opacities have similar values (same order of magnitude) around 0.001-0.01, and slightly higher values when considering large fractal particles. This is because the extinction factor of smaller particles is lower but is compensated by a larger mass of haze. These opacities are in the range of what has been estimated on Pluto, that is 0.003-0.012 \citep{Glad:16,Ster:15}, and thus suggest an acceptable order of magnitude for the mass of haze obtained. 
Comparing the haze distribution (obtained with and without South Pole N$_{2}$ condensation) with the observations (made by imaging with the instruments Ralph/MVIC and LORRI and by UV occultation with the Alice spectrometer) can help to reveal the presence or the absence of N$_{2}$ ice at the South Pole. A latitudinally homogeneous haze density with a slight peak above the North and particularly above the South Pole is typical of our simulation with South Pole N$_{2}$ condensation. Conversely, simulations without South Pole N$_{2}$ condensation show a more extensive haze in the summer hemisphere. 
Comparing the line of sight opacity profiles at the egress and the ingress points can also help to distinguish both cases. The opacity at the egress point is at least twice the opacity at the ingress point in the case without South Pole N$_{2}$ condensation, and no significant difference is obtained in the case without.
However, a latitudinally homogeneous haze density can also be the results of a long characteristic time for precursors growth (several terrestrial years), that allows precursors to be transported towards southern latitudes by radiative heating and meridional circulation. 
Finally, another way to distinguish both cases is to compare the haze distribution in the first kilometers above the surface. \autoref{hazesection_r10_cap} shows that the condensation flow induced by the presence of N$_{2}$ ice in the winter hemisphere leads to a lack of haze above the surface in the summer hemisphere, and an accumulation of haze between 3 and 20 kilometers in the winter hemisphere, which is more pronounced for small particle radii. 
Although the simulations were done with uniform particle sizes, in reality the haze particle size may be locally distributed and vary in space and time, especially in the vertical. Thus it may be more realistic to consider a distribution of haze particle sizes, in order to take into account the gravitational segregation.
Compared to the uniform size case, if 10 nm spherical particles in the upper atmosphere become fractal particles in the lower atmosphere, with same monomer radius, then there will be a change in opacity but not in haze vertical distribution (because the sedimentation velocity remains the same). If 10 nm spherical particles grow up to 100 nm during their fall down towards the surface, then the sedimentation velocity of the particle would change. The increase of the particle size during the fall would compensate the increase of atmospheric pressure and lead to a more homogeneous haze density with altitude. In addition, at the altitudes where transitions of particle size occur, layers of haze could form.



\ack
We wish to thank P. Rannou and E. Lellouch for helpful conversations and suggestions. We are grateful to P. Rannou for supplying a mean model of absorption and scattering by fractal particles of identical spheres. The authors thank the NASA \textit{New Horizons} team for their excellent work on a fantastic mission and their interest in this research.

\label{lastpage}

\bibliographystyle{plainnat}
\bibliography{bibliography}



\clearpage	

\clearpage


\end{document}